\documentclass[aps, prx, amssymb, longbibliography, twocolumn, superscriptaddress]{revtex4-2}
\usepackage[german,english]{babel}
\usepackage[utf8]{inputenc}

\usepackage{graphicx}
\usepackage{dcolumn}
\usepackage{bm}
\usepackage{physics}

\usepackage{dcolumn}
\usepackage{subfigure}
\usepackage{float}

\usepackage{tabularx}
\usepackage[normalem]{ulem}
\usepackage[usenames, dvipsnames]{color}
\usepackage{xcolor}

\usepackage[linesnumbered, ruled]{algorithm2e}

\usepackage{colortbl}
\usepackage{diagbox}
\usepackage[version=4]{mhchem}

\usepackage{theorem}

\usepackage{comment}

\usepackage{hyperref}
\hypersetup{                    
    colorlinks,
    citecolor=blue,
    filecolor=blue,
    linkcolor=blue,
    urlcolor=blue
}

\begin{document}

\title{Warm Start Adaptive-Bias Quantum Approximate Optimization Algorithm}

\author{Yunlong Yu}
\affiliation{ State Key Laboratory of Low Dimensional Quantum Physics, Department of Physics,\\Tsinghua University, Beijing 100084, China}

\author{Xiang-Bin Wang}
\affiliation{ State Key Laboratory of Low Dimensional Quantum Physics, Department of Physics,\\Tsinghua University, Beijing 100084, China}

\author{Nic Shannon}
\affiliation{Theory of Quantum Matter Unit, Okinawa Institute of Science and Technology Graduate University, Onna-son, Okinawa 904-0412, Japan}

\author{Robert Joynt}
\affiliation{Department of Physics, University of Wisconsin-Madison, Madison, WI 53706, USA}

\begin{abstract}
In the search for quantum advantage in real--world problems, one promising avenue is to use a quantum algorithm to improve on the solution found using an efficient classical algorithm. The quantum approximate optimization algorithm (QAOA) is particularly well adapted for such a "warm start" approach, and can be combined with the powerful classical Goemans-Williamson (GW) algorithms based on semi-definite programming. Nonetheless, the best way to leverage the power of the QAOA remains an open question. Here we propose a general model that describes a class of QAOA variants, and use it to explore routes to quantum advantages in a canonical optimization problem, MaxCut. 
For these algorithms we derive analytic expectation values of the cost Hamiltonian for the MaxCut problem in the level-1 case. Using these analytic results we obtain reliable averages over many instances for fairly large numbers of qubits. We find that the warm start adaptive-bias QAOA (WS-ab-QAOA) initialized by the GW algorithm outperforms previously proposed warm start variants on problems with $40$ to $180$ qubits. To assess whether a quantum advantage exists with this algorithm, we did numerical simulations with up to $1000$ qubits to see whether the level-1 WS-ab-QAOA can improve the GW solution for 3-regular graphs.  In fact the improvement in the $1000$-qubit case even in level 1 can only be matched by the GW algorithm after about $10^{5.5}$ random projections performed after the semi-definite program stage.  This work gives evidence that the final stage of optimization after an efficient classical algorithm has produced an approximate solution may be a place where quantum advantages can be realized.

\end{abstract}

\date{\today}
\maketitle

\section{Introduction}





 Quantum computing hardware continues to make strides~\cite{sycamore,zuchongzhi,zuchongzhi3}. Most recently, quantum error correction has been successfully demonstrated~\cite{qecc1,qecc2,qecc3,qecc4,qecc5,qecc6,qecc7}. However, large-scale fault-tolerant quantum computation~\cite{shor1,shor2} is still out of reach and we have not yet emerged from the age of noisy intermediate-scale quantum (NISQ) devices~\cite{nisq}. On this intermediate time scale, hybrid quantum-classical algorithms such as the variational quantum algorithm (VQA)~\cite{vqe,mcclean2016theory,vqa_review1,vqa_review2} are some of the most promising quantum algorithms to realize a quantum advantage.

For many paradigmatic combinatorial optimization problems, the leading candidate among hybrid algorithms is the quantum approximate optimization algorithm (QAOA).  It can be thought of as a kind of flexible Trotterization of a continuous quantum annealing process~\cite{qaoa_farhi,qaoa_lukin,qaoa_review1,qaoa_review2}. Extensive experimental demonstrations~\cite{sycamore_qaoa,Rydberg_qaoa,Rydberg_qaoa2,Rydberg_qaoa3} with up to 289 qubits using Rydberg atom arrays~\cite{Rydberg_qaoa} and numerical simulations for up to $54$ qubits with neural networks~\cite{qaoa_nn} have explored the possibility of a quantum advantage for QAOA.

The canonical benchmarking problem for QAOA in the literature is the MaxCut problem~\cite{qaoa_farhi,qaoa_lukin,qaoa_review1,qaoa_review2,qaoa_level23,abqaoa}, which aims to cut the vertex set of a graph into two subsets in such a way as to maximize the number of cuts, the edges connecting the two subsets. The performance criterion for an algorithm is usually the approximation ratio $r$~\cite{qaoa_farhi,qaoa_lukin}.  This is defined as the calculated cut value divided by the exact maximal cut value. It is NP-hard to obtain the solution with $r\geq 16/17 \approx 0.941$ on arbitrary  graphs~\cite{maxcut_hard}. 

The Goemans-Williamson (GW) algorithm is based on a semi-definite program (SDP)~\cite{SDP} followed by a random projection to produce an approximate solution. It solves the MaxCut problem in polynomial time with a guarantee that $r\geq 0.878$ for arbitrary graphs~\cite{GW}. There is evidence that this value cannot be substantially improved classically. Modifications to the SDP and the projection are not effective in doing so~\cite{bound}, and more generally, an improvement would contradict the Unique Games Conjecture ~\cite{conjecture,wsqaoa}. This possible gap between P and NP-hard is very suggestive: does there exist an intermediate $r$-value such that the problem is NP-intermediate, and thus analogous to factoring? Such problems are promising candidates for a quantum advantage. 

To find an approximate solution to the MaxCut problem or other combinatorial optimization problems through QAOA, the problem is converted into a cost Hamiltonian $H_\mathrm{C}$ of the Ising type, meaning that it contains only Pauli $Z$ operators.  Its ground state or the most highly excited state represents the solution~\cite{np_ising}. The evolution starts from the ground state of the mixing Hamiltonian $H_\mathrm{M}^\mathrm{s}$, which is usually the uniform superposition of Pauli $X$ operators of all the qubits. The unitary operators $\exp(-i\beta_l H_\mathrm{M}^\mathrm{s})$ and $\exp(-i\gamma_l H_\mathrm{C})$ are alternately applied $p$ times to produce the approximate solution.  $p$ is referred to the level. The schedule parameters $\Vec{\beta}$ and $\vec{\gamma}$ are trained to optimize the expectation values of the cost Hamiltonian~\cite{qaoa_farhi,qaoa_lukin}. Vector notation such as $\vec{\beta}$ and $\vec{\gamma}$ will be used to represent a set of parameters $\{\beta_1,\beta_2,...,\beta_p\}$  and  $\{\gamma_1,\gamma_2,...,\gamma_p\}$ henceforth.

To enhance the prospects for quantum advantage, numerous strategies have been proposed to further improve the QAOA, such as heuristic initialization methods of the schedule parameters~\cite{qaoa_lukin,tqa_qaoa}, modifications of the cost Hamiltonian~\cite{rqaoa,cvarqaoa} and the mixing Hamiltonian~\cite{abqaoa,abqaoa_sat,adapt_qaoa,wsqaoa}.  For the purposes of this paper, there are two types of modifications that are of most interest: inclusion of bias fields in the mixing Hamiltonian~\cite{abqaoa,abqaoa_sat}, and warm start, in which an approximate solution from a classical solver is used to determine the ansatz~\cite{wsqaoa,qaoa_warmest,qaoa_treepara,qaoa_stuck}. 

The adaptive-bias quantum approximate optimization algorithm (ab-QAOA) includes the adaptive longitudinal bias fields~\cite{bias_qaa} in the mixing Hamiltonian which are updated in the optimization process~\cite{abqaoa,abqaoa_sat}. The starting state is the ground state of the modified mixing Hamiltonian. For the MaxCut problem where the QAOA performs very well~\cite{qaoa_farhi,qaoa_lukin} and the SAT/Max-SAT problems where the QAOA appears to have difficulties~\cite{1_k_sat_qaoa,reachabiilty_deficits_sat}, the ab-QAOA has demonstrated a considerable speedup over the QAOA. The well-known easy-hard-easy phase transitions in the SAT problems~\cite{1_k_sat_qaoa} and the reachability-deficit phenomenon in the Max-SAT problems~\cite{reachabiilty_deficits_sat} are almost absent in the ab-QAOA.  

The recently proposed warm start QAOA variants modify the ansatz to encode the approximate solution from the classical solver~\cite{wsqaoa,qaoa_warmest,qaoa_treepara}. Additional Pauli $Y$, $Z$ operators operators are introduced in the mixing Hamiltonian and the starting state is accordingly adjusted either for recovering the input classical solution~\cite{wsqaoa} or satisfying the adiabaticity in the limit $p\rightarrow \infty$~\cite{qaoa_warmest}. The simulation on small-scale graphs suggests the superiority of warm start QAOA over the GW algorithm~\cite{wsqaoa,qaoa_warmest}. This motivates us to combine the current warm start strategy with the ab-QAOA to propose the warm start ab-QAOA (WS-ab-QAOA) to further improve the solutions and to test it on graphs with hundreds of vertices.

The WS-ab-QAOA differs from the ab-QAOA only in the inclusion of the classical solutions. Compared with other warm start variants, such as WS-QAOA~\cite{wsqaoa} and QAOA-warmest\cite{qaoa_warmest,qaoa_treepara}, the main feature of the WS-ab-QAOA is the update of the bias fields, which leads to a significant improvement on the input classical solution. 

Apart from the update of the bias fields, the WS-ab-QAOA is similar to the WS-QAOA~\cite{wsqaoa}. The encoding strategies of the classical solutions are the same: rotations around the $y$ axis are applied in order to encode the projected solutions in the classical procedure. A regularization parameter in WS-QAOA is adopted to avoid initialization at the poles on the Bloch sphere, which is unnecessary in the WS-ab-QAOA. Unlike the WS-ab-QAOA, the starting state in the WS-QAOA is not the ground state of the mixing Hamiltonian. For QAOA-warmest~\cite{qaoa_warmest}, the initialization strategy is different, since it is the solution before random projection that is used.  

In this paper, we propose a general model of the warm start QAOA variants, which can describe the WS-QAOA, the QAOA-warmest, the WS-ab-QAOA and other variants within a single framework. For the MaxCut problem on unweighted 3-regular (u$3$r) graphs, we derive analytical expressions for the expectation values of $H_\mathrm{C}$ in the level-1 case to use in an efficient numerical simulation. Based on this analytical method, we do numerical simulations of the quantum algorithms on systems with up to $180$ qubits. We show that when warm started from the GW solutions, the WS-ab-QAOA outperforms the WS-QAOA, the QAOA-warmest and even the input GW solutions in the level-1 case, where the WS-QAOA and QAOA-warmest are no better than the GW algorithm.

To further assess the possibility of a true quantum advantage over classical algorithms, we do numerical calculations on systems with up to $1000$ qubits on 3-regular graphs to compare WS-ab-QAOA with the GW algorithm, when multiple random projections are included to improve the solution. The multiple random projections mean running the GW algorithm multiple times since the SDP program is a convex optimization problem. On a system of $1000$ qubits it takes about $10^{5.5}$ projections for the GW algorithm to be competitive with the WS-ab-QAOA.

The paper is organized as follows: 
In Sec.~\ref{Sec:QAOA}, we give a brief introduction to the MaxCut problem and the quantum approximate optimization algorithm (QAOA).
In Sec.~\ref{Sec:Algorithm}, we propose the general model of warm start QAOA and show how to incorporate the various variants, such as WS-ab-QAOA, WS-QAOA and QAOA-warmest into this model. In addition, we prove that the level-1 WS-ab-QAOA can reproduce any input solution.  
In Sec.~\ref{Sec:Simulation Method},
we show how to compute the expectation value of the level-1 warm start QAOA variants efficiently on classical computers by an analytical method, which enables us to investigate many-qubit systems.  
In Sec.~\ref{Sec:Numerical Results}, we implement the WS-ab-QAOA numerically and address the issue of convergence. The performance of the WS-ab-QAOA is compared to other approaches.   
Sec.~\ref{Sec:Complexity} contains a complexity analysis of the classical simulation of the WS-ab-QAOA. 
Sec.~\ref{Sec:Conclusions} concludes with a summary of our findings and outlines potential future research directions.

\section{Quantum Approximate Optimization Algorithm}\label{Sec:QAOA}
\subsection{MaxCut problem}

The MaxCut problem is defined on a graph $G(\mathcal{V},\mathcal{E})$ where $\mathcal{V}$ is the vertex set and $\mathcal{E}$ is the edge set. In this paper, we consider only unweighted 3-regular (u$3$r) graphs: each vertex in the graph is connected to exactly $3$ other vertices and each edge is assigned the same weight $1$. The MaxCut problem is to cut the vertex set $\mathcal{V}$ into two subsets $\mathcal{V}_1$ and $\mathcal{V}_2$ such that the number of cuts (edges connecting $\mathcal{V}_1$ and $\mathcal{V}_2$) is maximized. 

One way to solve the MaxCut problem is to attach binary variables $x_j$ taking the values $\pm 1$ to each vertex. The two values indicate membership in $\mathcal{V}_1$ or $\mathcal{V}_2$. Hence the classical version of the cost function is 
\begin{align}
	C=\max_{x\in\{\pm1\}^n}\sum_{(j,k)\in \mathcal{E}}\frac{1}{4}(x_j-x_k)^2,\label{equ:maxcut_cost}
\end{align}
which means that if the edge $(j,k)$ is a cut then $x_j-x_k=\pm2$ and it will contribute $1$ to the cost function, while if the edge $(j,k)$ is not a cut then $x_j-x_k=0$ and it contributes $0$. The cost function $C$ represents the number of cuts. We will also refer to this number as the cut value.

The MaxCut problem can be approximately solved by the Goemans-Williamson (GW) algorithm~\cite{GW} based on a semi-definite program (SDP)~\cite{SDP} followed by random projections resulting in an approximate ratio of at least $0.878$ on all graphs. Thus the GW algorithm proceeds in 2 steps.  The solution space is first enlarged and an exact solution of an approximate version of the MaxCut problem is obtained. Then this solution is projected down to the allowed subset ($\pm 1$ on each vertex). The last step is done by random projections onto the hyperplane. A detailed description of the GW algorithm and how to implement it numerically can be found in Appendix~\ref{Appendix:GW}. 

The semi-definite program is technically efficient but rather time-consuming, so Burer and Monteiro propose to use a rank-$\kappa$ relaxation of $x$~\cite{BM}, which is denoted by rank-$\kappa$ Burer–Monteiro MaxCut ($\mathrm{BM}$-$\mathrm{MC}_\kappa$). The optimization of the cost function then becomes non-convex. More details can be found in Appendix~\ref{Appendix:BM}.
 
For quantum algorithms, it is convenient to indicate membership in $\mathcal{V}_1$ or $\mathcal{V}_2$ by the quantum state $|0\rangle$ or $|1\rangle$, which are eigenstates of the Pauli $Z$ operator. The MaxCut problem is equivalent to finding the ground state of the Hamiltonian
\begin{align}
	H_\mathrm{C}=\sum_{(j,k)\in \mathcal{E}} \frac{1}{2}(Z_j Z_k-1),
\end{align} 
where $Z_j$ is the Pauli $Z$ operator of the $j^\mathrm{th}$ qubit. The negative of the eigenvalues of $H_\mathrm{C}$ represent the cut values for the different spin configurations. Thus the negative of the expectation values of $H_\mathrm{C}$ can are the calculated cut values. 

The different algorithms yield rather different cut values. For a given graph, we denote the exact maximal cut value by $N_\mathrm{cut}$. The calculated cut values of the different algorithms are denoted by $N_\mathrm{cut}^\mathrm{alg}$, where the superscript $\mathrm{alg}$ represents the different solvers. Thus we have $N_\mathrm{cut}^\mathrm{GW}$ for the GW algorithm, $N_\mathrm{cut}^\mathrm{ab}$ for WS-ab-QAOA, $N_\mathrm{cut}^\mathrm{WS}$ for WS-QAOA, $N_\mathrm{cut}^\mathrm{warm}$ for QAOA-warmest and $N_\mathrm{cut}^\mathrm{s}$ for (cold-started) standard QAOA.

\subsection{QAOA}

The standard QAOA~\cite{qaoa_farhi} starts from the state $|\psi_0^\mathrm{s}\rangle$, the ground state of the mixing Hamiltonian $H_\mathrm{M}^\mathrm{s}$,
\begin{align}
	H_\mathrm{M}^\mathrm{s}=\sum_j X_j.
\end{align}
The unitary operators $\exp(-i \gamma_l H_\mathrm{C})$ and $\exp(-i \beta_l H_\mathrm{M}^\mathrm{s})$ are applied alternately $p$ times to generate the output state of the QAOA, where $p$ is the level, 
\begin{align}
		|\psi_{f}^\mathrm{s} \rangle=
	\prod_{l=1}^{p}
	\mathrm{e}^{-i\beta_{l}H_\mathrm{M}^\mathrm{s}} \mathrm{e}^{-i\gamma_{l}H_\mathrm{C}}|\psi_0^\mathrm{s}\rangle.\label{equ:sqaoa}
\end{align}

The parameters $(\vec{\gamma},\vec{\beta})$ are trained to make $\langle \psi_{f}^\mathrm{s}|H_\mathrm{C}| \psi_{f}^\mathrm{s} \rangle$, the expectation values of $H_\mathrm{C}$, as small as possible. $|\psi_{f}^\mathrm{s} \rangle$ can be understood as the result of an evolution from the ground state of $H_\mathrm{M}^\mathrm{s}$ to the ground state of $H_\mathrm{C}$.

\section{Warm Start QAOA}
\label{Sec:Algorithm}

In this section, we will give a unified model of warm start QAOA. WS-QAOA, QAOA-warmest and WS-ab-QAOA can all be described by this model. 

Let the input warm start state be $|\psi_{\mathrm{c}}\rangle$, which encodes the classical solution. A rotation operator $R_{{m_1}_j}^j(\delta_j)$ around the direction $\hat{m}_{1_j}$ is applied to qubit $j$ to obtain the starting state of the warm start QAOA,
\begin{align}
	|\psi_{0}\rangle=\tilde{R}_{m_1}(\vec{\delta})|\psi_\mathrm{c}\rangle,
\end{align}
where 
\begin{align}
    \tilde{R}_{m_1}(\vec{\delta})=\prod_j R_{{m_1}_j}^j(\delta_j).
\end{align}

The mixing Hamiltonian of the warm start QAOA can be regarded as the standard QAOA mixing Hamiltonian $H_\mathrm{M}^\mathrm{s}$ with a rotation $R_{{m_2}_j}^j(\alpha_j)$ to $X_j$ along the direction $\hat{m}_{2_j}$, 
\begin{align}
	H_\mathrm{M}=\sum_j R_{m_2}^j(\alpha_j) X_j R_{m_2}^j(-\alpha_j)=\tilde{R}_{m_2}(\vec{\alpha})H_\mathrm{M}^\mathrm{s} \tilde{R}_{m_2}(-\vec{\alpha}).
\end{align}
Note that $|\psi_{0}\rangle$ is not necessarily the ground state of $H_\mathrm{M}$. 

The output state of the warm start QAOA that approximates the ground state of $H_\mathrm{C}$ is,
\begin{align}
	|\psi_{f} \rangle=
	\prod_{l=1}^{p}
	\tilde{R}_{m_2}(\vec{\alpha})\mathrm{e}^{-i\beta_{l}H_\mathrm{M}^\mathrm{s}}\tilde{R}_{m_2}(-\vec{\alpha}) \mathrm{e}^{-i\gamma_{l}H_\mathrm{C}}\tilde{R}_{m_1}(\vec{\delta})|\psi_\mathrm{c}\rangle.\label{equ:general_ws}
\end{align}
This is a rather general expression. In this work, we mainly consider the case where $\hat{m}_{1_j}$ and $\hat{m}_{2_j}$ are both along the $y$ axis. The rotation operator is defined as
\begin{equation}
	 R_y^j(\alpha_j)=\mathrm{e}^{-i \alpha_j Y_j/2}. 
\end{equation} 

For the standard QAOA, the output state in Eq.~\eqref{equ:sqaoa} is given by taking each $\delta_j$ to be $-\pi/2$, $\alpha_j$ to be $0$ and $|\psi_{\mathrm{c}}\rangle$ to be $|0\rangle^{\otimes n}$ in Eq.~\eqref{equ:general_ws}.

\subsection{WS-ab-QAOA}\label{Sec:abqaoa}

In the ab-QAOA, the longitudinal adaptive bias fields $\vec{h}$ are incorporated in the mixing Hamiltonian~\cite{abqaoa,abqaoa_sat}.
The output state of the ab-QAOA is 
\begin{equation}
	\begin{split}
		|\psi_{f}^{\mathrm{ab}} \rangle&=
		\prod_{l=1}^{p}
		\mathrm{e}^{-i\beta_{l}H_\mathrm{M}^\mathrm{ab}} \mathrm{e}^{-i\gamma_{l}H_\mathrm{C}}|\psi_{0}^{\mathrm{ab}}\rangle,\\
		H_\mathrm{M}^\mathrm{ab}&=\sum_j\frac{X_j-h_jZ_j}{\sqrt{1+h_j^2}}\\
        &=\sum_j (\cos\alpha_j X_j-\sin\alpha_j Z_j)\\
        &=\tilde{R}_y(\vec{\alpha}) H_\mathrm{M}^\mathrm{s} \tilde{R}_y(-\vec{\alpha}),
        \end{split}
\end{equation}
where $|\psi_{0}^{\mathrm{ab}}\rangle$ is the corresponding ground state of $H_\mathrm{M}^\mathrm{ab}$. The angle $\alpha_j$ defined by 
\begin{equation}
	\begin{split}
		\cos \alpha_j&=\frac{1}{\sqrt{1+h_j^2}}, \\
		\sin \alpha_j&=\frac{h_j}{\sqrt{1+h_j^2}}.
	\end{split}
\end{equation} 
$|\psi_{0}^{\mathrm{ab}}\rangle$ can be represented by the rotation of $|\psi_{0}^{\mathrm{s}}\rangle$, 
\begin{align}
	|\psi_{0}^{\mathrm{ab}}\rangle=\tilde{R}_y(\vec{\alpha}) |\psi_{0}^{\mathrm{s}}\rangle=\tilde{R}_y(\vec{\alpha}-\frac{\pi}{2}) |0\rangle^{\otimes n}.
\end{align}

The bias field parameters $\vec{h}$ are updated based on the expectation values of the $Z_j$ operators in addition to the optimization of $(\vec{\gamma},\vec{\beta})$. The basic idea of the ab-QAOA, as explained in Refs.~\cite{abqaoa,abqaoa_sat}, is that the bias field parameter $\vec{h}$ encodes more and more solution information during the optimization. This idea carries over to the WS-ab-QAOA. In ab-QAOA, the bias field parameters are updated according to the rule,
\begin{align}
	h_j=h_j-\ell(h_j-\langle Z_j \rangle).
\end{align} 
This is because a bias to the state $|0(1)\rangle_j$ is expected by setting $h_j=1(-1)$~\cite{abqaoa,abqaoa_sat}. A slightly modified update rule will be adopted in this paper as explained below.

The classical solution $|\psi_{\mathrm{c}}\rangle$ from the GW algorithm is used to warm start the WS-ab-QAOA. It is a computational basis state with the state of the $j^\mathrm{th}$ qubit is $|\psi_{\mathrm{c}}\rangle_j=|0(1)\rangle_j$. It can be generated by the application of the rotation around the $y$ axis
\begin{align}
    |\psi_{\mathrm{c}}\rangle=\tilde{R}_y(\vec{\alpha}^0-\frac{\pi}{2}) |0\rangle^{\otimes n},
\end{align}
where $\alpha_{j}^0=\pi/2 (-\pi/2)$ corresponding to $|\psi_\mathrm{c}\rangle_j=|0(1)\rangle_j$. Setting $\vec{\delta}=\vec{\alpha}-\vec{\alpha}^0$, the output state of the WS-ab-QAOA is
\begin{equation}
	|\psi_{f}^{\mathrm{ab}} \rangle=
	\prod_{l=1}^{p}
\tilde{R}_{y}(\vec{\alpha})	\mathrm{e}^{-i\beta_{l}H_\mathrm{M}^\mathrm{s}} \tilde{R}_{y}(-\vec{\alpha})	\mathrm{e}^{-i\gamma_{l}H_\mathrm{C}}\tilde{R}_{y}(\vec{\delta})	|\psi_{\mathrm{c}}\rangle,\label{equ:general_ab}
\end{equation}
which is the special case of Eq.~\eqref{equ:general_ws}.

We now show that for the unweighted MaxCut problem on odd-degree graphs (u$\mathcal{R}$r graphs with $\mathcal{R}$ odd), the level-$1$ WS-ab-QAOA can recover the input warm start state $|\psi_\mathrm{c}\rangle$. Note that the $|\psi_\mathrm{c}\rangle$ is always a classical product state.  Thus we need only to show that $|\psi_{f}^{\mathrm{ab}} \rangle$ with $p=1$ can be decomposed into the tensor product state and state of the $j^\mathrm{th}$ qubit can be either $|0\rangle_j$ or $|1\rangle_j$ depending on the input $h_j$.    

To do this set $\gamma_{1}=\pi$, giving
\begin{align}
	\begin{aligned}
		\mathrm{e}^{-i\gamma_{1}H_\mathrm{C}}&=\prod_{(j,k)\in \mathcal{E}} i (\cos\frac{\pi}{2}\mathcal{I}-i\sin\frac{\pi}{2}Z_jZ_k)\\
		&=\prod_{j}Z_j,
	\end{aligned}
\end{align}
where we drop the global phase. This simple form comes from the fact that each $Z_j$ appears $\mathcal{R}$ times in the product $\prod_{jk}Z_jZ_k$, so only $\prod_{j}Z_j $ is left. This also implies that for fixed $\gamma_{1}=\pi$, there is no entanglement in $|\psi_{f}^{\mathrm{ab}} \rangle$, so $|\psi_{f}^{\mathrm{ab}} \rangle$ is a classical product state.

For the state of the $j^\mathrm{th}$ qubit, we have 
\begin{align}
	|\psi_{f}^{\mathrm{ab}} \rangle_j=R_y^j(\alpha_j) \mathrm{e}^{-i\beta_{1}X_j} R_y^j(-\alpha_j) Z_j R_y^j(\alpha_j-\frac{\pi}{2})|0\rangle_j, 
\end{align} 
where the global phase is again ignored. Setting $\beta_1=\pi/2$ we find
\begin{align}
    \begin{aligned}
    	|\psi_{f}^{\mathrm{ab}} \rangle_j&=R_y^j(\alpha_j) X_j R_y^j(-\alpha_j) Z_j R_y^j(\alpha_j-\frac{\pi}{2})|0\rangle_j\\
        &=\left(\begin{array}{c}
		\cos\frac{1}{4}(\pi+6\alpha_j) \\
		\sin\frac{1}{4}(\pi+6\alpha_j) 
	\end{array}\right)_j.
    \end{aligned}\label{equ:ab_product}
\end{align} 
When $\alpha_j=-\pi/6$, $|\psi_{f}^{\mathrm{ab}} \rangle_j=|0\rangle$ , $h_j=-\sqrt{3}/3$ and $\alpha_j=\pi/6$, $|\psi_{f}^{\mathrm{ab}} \rangle_j=|1\rangle$ , $h_j=\sqrt{3}/3$. This completes the proof. This property of the WS-ab-QAOA motivates us to modify the update rule to a function that depends on $ h_j+\frac{\sqrt{3}}{3}\langle Z_j\rangle$.

We take $|\psi_\mathrm{c}\rangle$ from the GW algorithm after the random projection procedure and the corresponding cut value is denoted by $N_\mathrm{cut}^\mathrm{GW}$. The cut value of the ab-QAOA from the warm start classical solution is denoted by $N_\mathrm{cut}^\mathrm{ab}$, according to the fact above, we have
\begin{align}
	N_\mathrm{cut}^\mathrm{ab}\geq N_\mathrm{cut}^\mathrm{GW},
\end{align}
which means the ab-QAOA is not worse than the GW algorithm when warm started from the GW solution. 

In Appendix~\ref{Appendix:abQAOA}, based on this performance guarantee, we prove that when the bias-field update is turned off, in the special case where $\gamma_1=0$ or $\vec{\delta}=0$, the WS-ab-QAOA can not produce a better solution than the input solution, which emphasizes the importance for updating the bias fields. 

If $\mathcal{R}$ is even, then when $\gamma_{1}=\pi$,
\begin{align}
	\begin{aligned}
		\mathrm{e}^{-i\gamma_{1}H_{C}}&=\mathcal{I},\\
		|\psi_{f}^{\mathrm{ab}} \rangle_j&=R_y^j(\alpha_j) \mathrm{e}^{-i\beta_{1}X_j} R_y^j(-\alpha_j) R_y^j(\alpha_j-\frac{\pi}{2})|0\rangle_j\\
        &=\left(\begin{array}{c}
			\cos\frac{1}{2}(\alpha_j-\frac{\pi}{2}) \\
			\sin\frac{1}{2}(\alpha_j-\frac{\pi}{2}) 
		\end{array}\right)_j,
	\end{aligned}
\end{align}
where the global phase is ignored. It is obvious that $|\psi_{f}^{\mathrm{ab}} \rangle_j$ can not be $|0(1)\rangle_j$ since $\alpha_j \neq \pi/2$ and $\cos\alpha_j\neq 0$. Thus the proof does not go through for this case.

\subsection{WS-QAOA}\label{sec:wsqaoa}
The WS-QAOA differs from the WS-ab-QAOA by inverting the sign of the off-diagonal elements in the mixing Hamiltonian~\cite{wsqaoa},
\begin{align}
	\begin{split}
		H_\mathrm{M}^\mathrm{ws}&=\sum_j (-\cos\theta_j X_j-\sin\theta_j Z_j)\\
        &=\tilde{R}_y(\pi-\vec{\theta}) H_\mathrm{M}^\mathrm{s} \tilde{R}_y(\vec{\theta}-\pi),
	\end{split}
\end{align}
where $\theta_j=2\arcsin(\sqrt{c_j})-\pi/2$ and $c_j\in[0,1]$ encodes the input solution. The starting state is, 
\begin{align}
	|\psi_{0}^{\mathrm{ws}}\rangle=\tilde{R}_y(\vec{\theta}) |\psi_{0}^{\mathrm{s}}\rangle.
\end{align}
To avoid initializing the state to be exactly $|0\rangle$ or $|1\rangle$ when $c_j=0$ or $1$ where only Pauli $Z$ terms appear in $H_\mathrm{M}^\mathrm{ws}$, the following strategy with regularization parameter $\epsilon$ is adopted to create a superposition state:
\begin{align}
	\theta_j=\left\{\begin{array}{ll}
		2\arcsin(\sqrt{c_j})-\pi/2, & c_j\in[\epsilon,1-\epsilon],\\
		2\arcsin(\sqrt{\epsilon})-\pi/2, & c_j\leq\epsilon,\\
		2\arcsin(\sqrt{1-\epsilon})-\pi/2, & c_j\geq1-\epsilon.\\
	\end{array}\right.
\end{align}

The output state of the WS-QAOA is
\begin{align}
    \begin{aligned}
    	|\psi_{f}^{\mathrm{WS}} \rangle=
			\prod_{l=1}^{p}&
			\tilde{R}_y(\pi-\vec{\theta})\mathrm{e}^{-i\beta_{l}H_\mathrm{M}^\mathrm{s}}\tilde{R}_y(\vec{\theta}-\pi) \mathrm{e}^{-i\gamma_{l}H_\mathrm{C}}\\
            &\tilde{R}_y(\vec{\theta}-\frac{\pi}{2}) |0\rangle^{\otimes n}.
    \end{aligned}
\end{align} 
Just as for the WS-ab-QAOA, we can set $\alpha_j^0=\pi/2(-\pi/2)$ to recover the warm
start state $|\psi_{\mathrm{c}}\rangle$ by a single $\tilde{R}_y$ rotation. This gives 
\begin{equation}
    \begin{split}
    	|\psi_{f}^{\mathrm{WS}} \rangle=\prod_{l=1}^p&
    \tilde{R}_y(\pi-\vec{\theta})\mathrm{e}^{-i\beta_{l}H_\mathrm{M}^\mathrm{s}}\tilde{R}_y(\vec{\theta}-\pi)\mathrm{e}^{-i\gamma_{l}H_\mathrm{C}}\\
    &\tilde{R}_{y}(\vec{\theta}-\vec{\alpha}^0)	|\psi_{\mathrm{c}}\rangle.
    \end{split}
\end{equation}
Denoting $\vec{\alpha}=\pi-\vec{\theta}$ and $\vec{\delta}=\pi-\vec{\alpha}-\vec{\alpha}^0$, we have
\begin{equation}
	|\psi_{f}^{\mathrm{WS}} \rangle=
	\prod_{k=1}^{p}
	\tilde{R}_{y}(\vec{\alpha})	\mathrm{e}^{-i\beta_{k}H_\mathrm{M}^\mathrm{s}} \tilde{R}_{y}(-\vec{\alpha})	\mathrm{e}^{-i\gamma_{k}H_\mathrm{C}}\tilde{R}_{y}(\vec{\delta})	|\psi_{\mathrm{c}}\rangle,
\end{equation}
which is the special case of Eq.~\eqref{equ:general_ws}.

Note that in Ref.~\cite{wsqaoa}, the mixing Hamiltonian of the standard QAOA is $-\sum_j X_j$, from which the WS-QAOA is constructed. In this paper, the mixing Hamiltonian is $\sum_j X_j$, based on which we derive a slightly different version of WS-QAOA. The level-1 WS-QAOA is also able to recover the input warm start state as shown in Appendix~\ref{Appendix:WSQAOA}.

\subsection{QAOA-warmest}
For QAOA-warmest, the mixing Hamiltonian is
\begin{align}
	\begin{split}
		H_\mathrm{M}^\mathrm{warm}&=-\sum_j (c_{x_j} X_j+c_{y_j} Y_j+c_{z_j} Z_j),\\
		c_{x_j}&=\sin\theta_j\cos\phi_j,\\
		c_{y_j}&=\sin\theta_j\sin\phi_j,\\
		c_{z_j}&=\cos\theta_j,
	\end{split}\label{equ:mixing_warm}
\end{align}
and the corresponding starting ground state is,
 \begin{align}
 	\begin{split}
 		|\psi_0^\mathrm{warm}\rangle&=\bigotimes_j \left(\cos\frac{\theta_j}{2}|0 \rangle+\mathrm{e}^{i\phi_j}\sin\frac{\theta_j}{2}|1\rangle\right).
 	\end{split}\label{equ:psi0_warmest}
 \end{align}
Here the angles $(\vec{\theta},\vec{\phi})$ encode the input warm start state and can be obtained from the classical $\mathrm{BM}$-$\mathrm{MC}_\kappa$ algorithm with the classical cost function 
\begin{align}
	\begin{aligned}
		\mathrm{maximize} &\sum_{(j,k)\in \mathcal{E}}\frac{1}{4}(x_j-x_k)^2,\\
		\mathrm{s.t.}\quad & ||x_j||=1,\\
		& x_j \in \mathbf{R}^\kappa.
	\end{aligned}\nonumber
\end{align} 
Here each $x_j$ is a relaxed $\kappa$-dimensional vector instead of the $n$-dimensional vector in Eq.~\eqref{equ:maxcut_cost}. 

Note that in Ref.~\cite{qaoa_warmest}, the evolution from the most excited state to the most excited state is considered, while in this paper, the evolution is from the ground state to the ground state. The two approaches differ only in the sign of $\vec{\beta}$. A detailed description of the QAOA-warmest and the performance guarantee when warm started from GW solution can be found in Appendix~\ref{Appendix:QAOAwarmest}.

Note that in Ref.~\cite{qaoa_warmest}, $\kappa=2$ QAOA-warmest is no worse than $\kappa=3$ but has a much easier implementation. So we mainly consider $\kappa=2$ in this paper, where $\phi_j=0$, replace $\theta_j$ with $\pi/2-\theta_j$, then $c_{x_j}=\cos\theta_j$ and $c_{z_j}=\sin\theta_j$. For the classical optimization, each $x_j$ can be represented by a point on a radius-1 circle, $x_j=(\cos\theta_j,\sin\theta_j)$. The optimization is done over $\vec{\theta}$. After getting the optimal $\vec{\theta}^{\mathrm{opt}}$ from  $\mathrm{BM}$-$\mathrm{MC}_2$ algorithm, apply the "vertex-at-top" strategy to obtain the input $\vec{\theta}$ of the QAOA-warmest, which means randomly selecting one component in $\vec{\theta}^{\mathrm{opt}}$ and then shifting all the angles by this value~\cite{qaoa_warmest_rotation}. 
Defining $\theta_j=\alpha_j-\pi/2$, we have 

\begin{align}
	\begin{split}
		H_\mathrm{M}^\mathrm{warm}&=\sum_j (\cos\alpha_jX_j-\sin\alpha_j Z_j),\\
		|\psi_0^\mathrm{warm}\rangle&=\bigotimes_j \left[\sin(\frac{\alpha_j}{2}+\frac{\pi}{4})|0 \rangle_j-\cos(\frac{\alpha_j}{2}+\frac{\pi}{4})|1\rangle_j\right]\\
        &=\tilde{R}_y(\vec{\alpha}-\frac{\pi}{2}) |0\rangle^{\otimes n}.
	\end{split}
\end{align}

Note that the initial state $|\psi_0^\mathrm{warm}\rangle$ is not a computational basis state, in apparent contradiction to Eq.~\eqref{equ:general_ws}. However, if we set $|\psi_\mathrm{c}\rangle=|0\rangle^{\otimes n}$ and $\vec{\delta}=\vec{\alpha}-\pi/2=\vec{\theta}$, the output state of the QAOA-warmest is still
\begin{equation}
	|\psi_{f}^{\mathrm{warm}} \rangle=
	\prod_{l=1}^{p}
	\tilde{R}_{y}(\vec{\alpha})	\mathrm{e}^{-i\beta_{l}H_\mathrm{M}^\mathrm{s}} \tilde{R}_{y}(-\vec{\alpha})	\mathrm{e}^{-i\gamma_{l}H_\mathrm{C}}\tilde{R}_y(\vec{\delta})|\psi_{\mathrm{c}}\rangle,
\end{equation} 
which is exactly the ab-QAOA relationship. We emphasize that the differences are the encoding strategy of the classical solution and the update of the bias fields. The solution is encoded in $\vec{\theta}$ instead of $|\psi_\mathrm{c}\rangle$.

\section{Simulation Method} \label{Sec:Simulation Method}

In the level-1 case, we can calculate the expectation value of the cost Hamiltonian efficiently on classical computers for the u$3$r MaxCut problems.  This allows us to obtain very high-quality numerical data.  We can compute the expectation value but not the final output state. This is sufficient for our purposes.  

\begin{figure}[ht]
	\centering
	\includegraphics[scale=0.3]{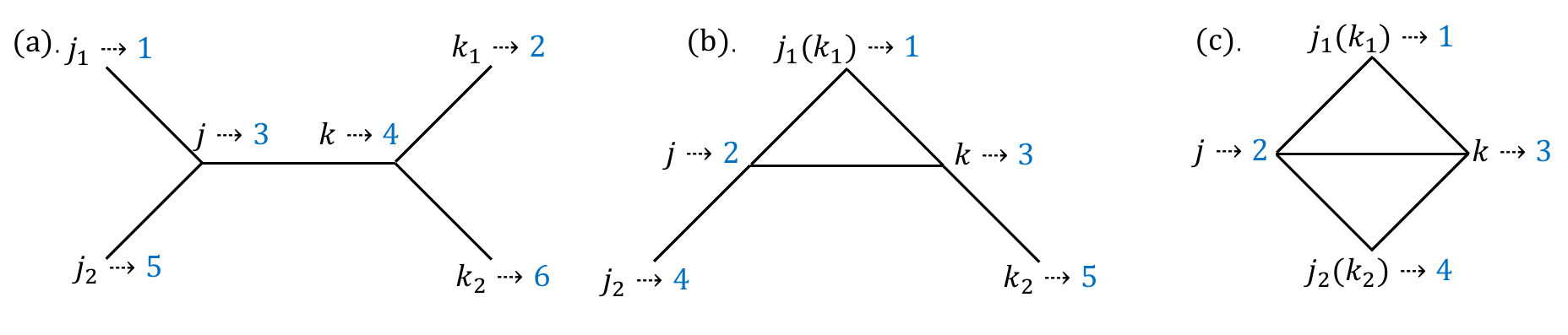} 
	\caption{Subgraphs considered in the level-1 case for the u$3$r graphs. The letters $j,k,j_1,j_2,k_1,k_2$ represent the order in the original graphs while the numbers $1,2,3,4,5,6$ represent the order in the reduced subgraphs.}\label{fig:subgraph}
\end{figure}

The expectation value of the cost Hamiltonian in the output state of warm-started QAOA is 
\begin{align}
	\begin{split}
	\langle H_\mathrm{C}\rangle=&\langle \psi_f^\mathrm{ab} |H_\mathrm{C}| \psi_f^\mathrm{ab}\rangle \\
        =&\sum_{(j,k)\in \mathcal{E}} \langle \psi_{\mathrm{c}}|\tilde{R}_y(-\vec{\delta})\biggl[\mathrm{e}^{i\gamma_{1}H_\mathrm{C}}\tilde{R}_{y}(\vec{\alpha})	\mathrm{e}^{i\beta_{k}H_\mathrm{M}^\mathrm{s}}\tilde{R}_{y}(-\vec{\alpha}) \\
	&\frac{Z_jZ_k}{2}\tilde{R}_{y}(\vec{\alpha})	\mathrm{e}^{-i\beta_{k}H_\mathrm{M}^\mathrm{s}} \tilde{R}_{y}(-\vec{\alpha})        \mathrm{e}^{-i\gamma_{1}H_\mathrm{C}}-\frac{1}{2}\biggr]\\
        &\tilde{R}_y(\vec{\delta})|\psi_{\mathrm{c}}\rangle.
	\end{split}\nonumber
\end{align}
Considering a single edge $(j,k)$, in the corresponding unitary operator $\exp(-i\beta_{1}H_\mathrm{M}^\mathrm{s})$, only the $j^\mathrm{th}$ and $k^\mathrm{th}$ qubits will contribute to the expectation value of this edge, while in $\exp(-i\gamma_{1}H_\mathrm{C})$, only the edges connected to the qubit $j$ and the edges connected to the qubit $k$ will contribute to the final expectation value of this edge. Thus for u$3$r graphs we have,
\begin{align}
	\begin{aligned}
		\langle H_\mathrm{C}\rangle=&\sum_{\substack{{ (j,k)}\in \mathcal{E}\\(j,j_1)\in \mathcal{E},j_1\neq k\\
				(k,k_1) \in \mathcal{E}, k_1 \neq j }} \langle\psi_{\mathrm{c}}|\tilde{R}_y(-\vec{\delta})\biggl[\mathrm{e}^{i\gamma_{1} (Z_j Z_{j_1}+Z_k Z_{k_1}+Z_j Z_{k})/2}\\
		&R_y^j(\alpha_j)R_y^k(\alpha_k)\mathrm{e}^{i\beta_{1}(X_j+X_k)}R_y^k(-\alpha_k)R_y^j(-\alpha_j)\frac{Z_jZ_k}{2}\\
		&R_y^j(\alpha_j)R_y^k(\alpha_k)\mathrm{e}^{-i\beta_{1}(X_j+X_k)}R_y^k(-\alpha_k)R_y^j(-\alpha_j)\\
		&\mathrm{e}^{-i\gamma_{1} (Z_j Z_{j_1}+Z_k Z_{k_1}+Z_j Z_{k})/2}-\frac{1}{2}\biggr]\tilde{R}_y(\vec{\delta})|\psi_{\mathrm{c}}\rangle.
	\end{aligned}\label{equ:zz}
\end{align}
where the qubits $j_1$ and $k_1$ can be equivalent. Specifically, we only need to consider the subgraphs with $4$,$5$ and $6$ qubits as shown in Fig~\ref{fig:subgraph}. 

To evaluate Eq.~\eqref{equ:zz}, the following rotation of $Z_j$ is necessary:
\begin{align}
	\begin{aligned}
		&R_y^j(\alpha_{j}) R_x^j(-2\beta_1) R_y^j(-\alpha_{j}) Z_j R_y^j(\alpha_{j}) R_x^j(2\beta_1) R_y^j(-\alpha_{j})\\
	=&C_j^Z Z_j+C_j^Y Y_j + C_j^X X_j,
	\end{aligned}
\end{align} 
where 
\begin{align}
    \begin{aligned}
        &C_j^Z=(1-2\cos^2\alpha_{j}\sin^2\beta_{1}),\\
        &C_j^Y=\cos\alpha_{j}\sin2\beta_1,\\
        &C_j^X=-\sin2\alpha_{j}\sin^2\beta_1.
    \end{aligned}
\end{align}
Finally we have 
\begin{align}
	\begin{aligned}
		&\langle H_\mathrm{C}\rangle\\
        =&\sum_{(j,k)\in \mathcal{E}} \langle\psi_{\mathrm{c}}|\tilde{R}_y(-\vec{\delta})\bigg[\mathrm{e}^{i\gamma_{1} H_\mathrm{C}}
		(C_j^Z Z_j+C_j^Y Y_j+C_j^X X_j)\\
		&(C_k^Z Z_k+C_k^Y Y_k + C_k^X X_k)
		\mathrm{e}^{-i\gamma_{1} H_\mathrm{C}}-\frac{1}{2}\biggr]\tilde{R}_y(\vec{\delta})|\psi_{\mathrm{c}}\rangle\\
		=&\sum_{(j,k)\in \mathcal{E}}\biggl(C_j^ZC_k^Z E_{Z_jZ_k}+C_j^ZC_k^Y E_{Z_jY_k}+C_j^Y C_k^Z E_{Y_jZ_k}\\
		&+C_j^ZC_k^X E_{Z_jX_k}+C_j^X C_k^Z E_{X_jZ_k}+C_j^Y C_k^Y E_{Y_jY_k}\\
        &+C_j^X C_k^X E_{X_jX_k}+C_j^Y C_k^X E_{Y_jX_k}+C_j^X C_k^Y E_{X_jY_k}-\frac{1}{2}\biggr),
	\end{aligned}
\end{align}
where the expectation values of the two-body Pauli operators such as $E_{Z_jZ_k}$ are defined by
\begin{align}
    E_{Z_jZ_k}=\langle\psi_{\mathrm{c}}|\tilde{R}_y(-\vec{\delta}) \mathrm{e}^{i\gamma_{1} H_\mathrm{C}}Z_jZ_k\mathrm{e}^{-i\gamma_{1} H_\mathrm{C}}\tilde{R}_y(\vec{\delta}) |\psi_{\mathrm{c}}\rangle.
\end{align}
In Appendix~\ref{Appendix:Analytical}, we will show how to calculate them and present the analytical results for u$3$r graphs.

The expectation values $\{\langle Z_j \rangle\}$ are also needed, and can be obtained by similar methods. Only the vertices connected to vertex $j$ are required in the calculation:
\begin{align}
	\begin{aligned}
		&\langle Z_j \rangle\\
        =&\sum_{k\,\mathrm{s.t.}\, (j,k)\in \mathcal{E}} \langle\psi_{\mathrm{c}}|\tilde{R}_y(-\vec{\delta})\biggl[\mathrm{e}^{i\gamma_{1} Z_j Z_{k}/2}R_y^j(\alpha_j)\mathrm{e}^{i\beta_{1}X_j}R_y^j(-\alpha_j)\\
		& Z_j R_y^j(\alpha_j)\mathrm{e}^{-i\beta_{1}X_j}R_y^j(-\alpha_j)
		\mathrm{e}^{-i\gamma_{1} Z_j Z_{k}/2}\biggr]\tilde{R}_y(\vec{\delta})|\psi_{\mathrm{c}}\rangle.
	\end{aligned}\label{equ:z}
\end{align}
By a similar process,
\begin{align}
	\begin{aligned}
		\langle Z_j\rangle
        =&\sum_{k\,\mathrm{s.t.}\, (j,k)\in \mathcal{E}}\langle\psi_{\mathrm{c}}|\tilde{R}_y(-\vec{\delta})\mathrm{e}^{i\gamma_{1} H_\mathrm{C}}
		\biggl(C_j^Z Z_j+C_j^Y Y_j\\
		& + C_j^X X_j\biggr)\mathrm{e}^{-i\gamma_{1} H_\mathrm{C}}\tilde{R}_y(\vec{\delta})|\psi_{\mathrm{c}}\rangle\\
		=&\sum_{k\,\mathrm{s.t.}\, (j,k)\in \mathcal{E}}
		(C_j^Z E_{Z_j}+C_j^Y E_{Y_j} + C_j^X E_{X_j}),
	\end{aligned}
\end{align}
with the single-body expectation values such as $E_{Z_j}$ defined as
\begin{align}
    E_{Z_j}=\langle\psi_{\mathrm{c}}|\tilde{R}_y(-\vec{\delta}) \mathrm{e}^{i\gamma_{1} H_\mathrm{C}}Z_j\mathrm{e}^{-i\gamma_{1} H_\mathrm{C}}\tilde{R}_y(\vec{\delta}) |\psi_{\mathrm{c}}\rangle.
\end{align}

Further analytical results for the single-body Pauli operators can be found in Appendix~\ref{Appendix:Analytical}. Since there are at most $6$ vertices in the subgraphs, brute-force numerical computation of Eq.~\eqref{equ:zz} and Eq.~\eqref{equ:z} is also possible. However, we note that the analytical method is faster.

\section{Numerical Results}\label{Sec:Numerical Results}
\subsection{Comparison of Different Warm Start Methods}
In this section, we compare numerically the performances of level-1 WS-QAOA, QAOA-warmest and WS-ab-QAOA for the MaxCut problems on u3r graphs.   

For level-1 WS-ab-QAOA, as mentioned in Sec.~\ref{Sec:abqaoa}, when $\gamma_1=\pi,\beta_1=\pi/2,h_j=\pm \sqrt{3}/3$ for all $h_j$, the output state is a product state with $\langle Z_j\rangle=\mp1$. This motivates us to regard $h_j+\frac{\sqrt{3}}{3}\langle Z_j\rangle$ as the gradient of $
h_j$. When $h_j=1/\sqrt{3}$ or $-1/\sqrt{3}$ and the state of the $j^\mathrm{th}$ qubit is $|1(0)\rangle_j$, the gradient of $h_j$ is $0$. That's where the update of $h_j$ stops as used in the conventional gradient descent algorithm. Thus the convergence of the bias field parameters lead to a computational basis state. 

Borrowing an idea from the Adam stochastic gradient descent algorithm~\cite{adam}, which is a first-order gradient-based optimization method, we define the gradient of $h_j$ in $t^\mathrm{th}$ iteration as 
\begin{align}
    g_{h_j}^{(t)}=h_j^{(t)}+\frac{\sqrt{3}}{3}\langle Z_j\rangle^{(t)},
\end{align}
where the superscript $(t)$ means the $t^\mathrm{th}$ iteration.
The update of $h_j$ is based on its  $1^\mathrm{st}$ moment estimate $m_{h_j}$ and $2^\mathrm{nd}$ raw moment estimate $v_{h_j}$ defined by 
\begin{align}
	\begin{aligned}
		m_{h_j}^{(t)}&= b_1 m_{h_j}^{(t-1)}+(1-b_1)g_{h_j}^{(t)},\\
		v_{h_j}^{(t)}&= b_2 v_{h_j}^{(t-1)} +(1-b_2)g_{h_j}^{(t)2}.\\
	\end{aligned}
\end{align}
The parameters $b_1$ and $b_2$ are taken to be  $b_1=0.9$ and $b_2=0.999$, which are the exponential decay rates for the moment estimates. If the gradient parameter $g_{h_j}$ is near $0$, $m_{h_j}$ and $v_{h_j}$ will decay to $0$ exponentially. The update rule of $h_j$ in $t^\mathrm{th}$ iteration with adaptive learning rate $\ell_t$ is
\begin{align}
	\begin{aligned}
		h_j^{(t)}&= h_j^{(t-1)}-\ell_t \frac{m_{h_j}^{(t)}}{\sqrt{v_{h_j}^{(t)}}+\varepsilon},\\
        \ell_t&=\ell_0 \frac{\sqrt{1-b_2^t}}{(1-b_1^t)}.
	\end{aligned}
\end{align}
where $\varepsilon$ is taken to be $\varepsilon=10^{-8}$ to avoid zero denominators. The initial learning rate is set to be $\ell_0=0.4$. A zero-gradient leads to the convergence of $h_j$.

For the ab-QAOA the bias field parameters $\vec{h}$ encode the approximate solution to the MaxCut problem~\cite{abqaoa_sat}.  This is formalized by constructing the bias state$|\psi_{\mathrm{bias}}\rangle$. This is done by mapping the final bias fields $h_j>0$ to $|1\rangle_j$ and $h_j<0$ to $|0\rangle_j$ as a result of the correspondence above. We will call this the bias-state method henceforth. As shown in Appendix~\ref{Sec:diff}, the $|\psi_{f}^{\mathrm{ab}}\rangle$ converges to $|\psi_{\mathrm{bias}}\rangle$ in all cases. It is slightly simpler to use this state instead of $|\psi_{f}^{\mathrm{ab}}\rangle$ and this is what we do here. Besides, the partition of the vertices can be obtained from the bias-state method. The calculated cut values from the bias-state method are denoted by
$N_\mathrm{cut}^\mathrm{bias}$.

\begin{figure}[ht]
	\centering
	\includegraphics[scale=0.45]{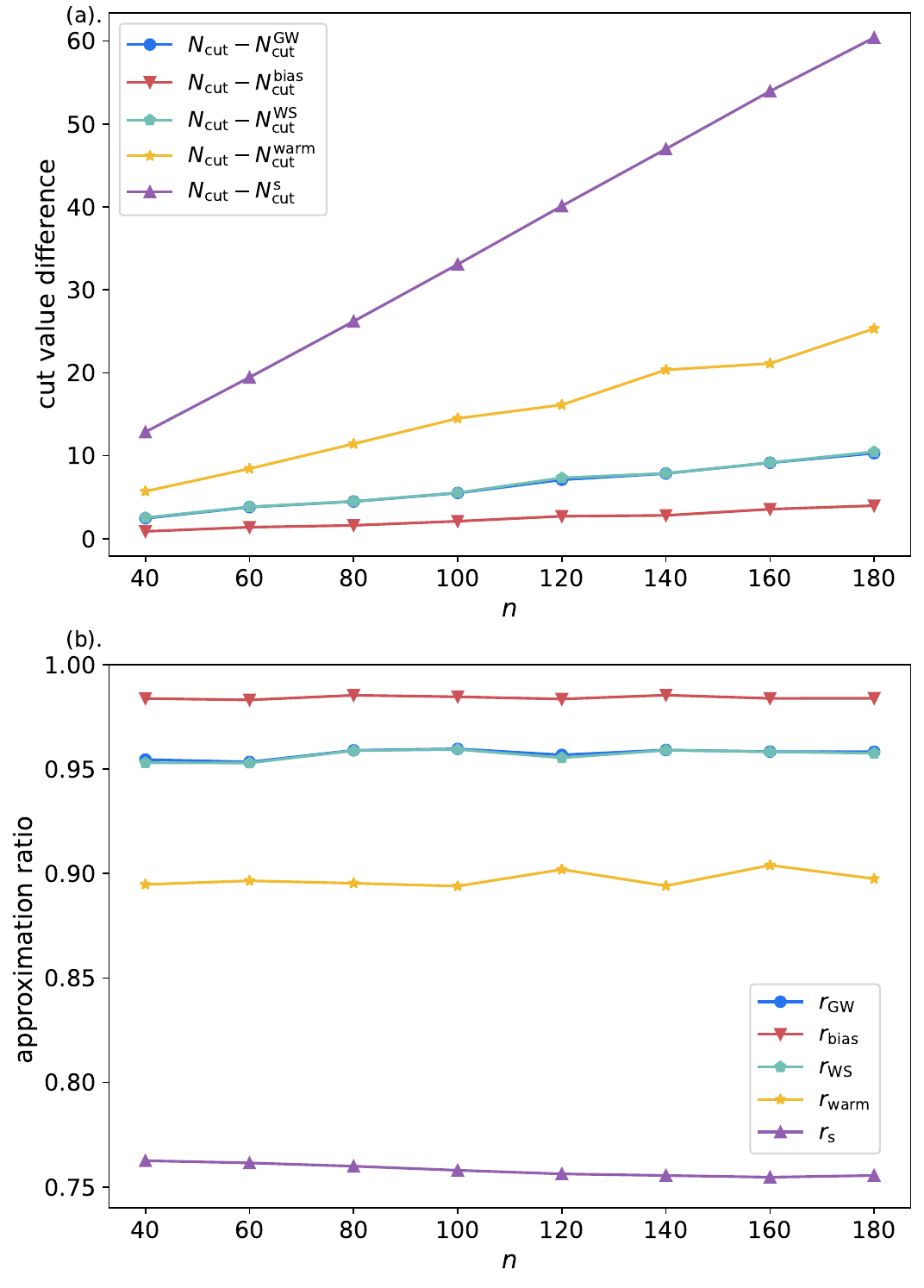} 
	\caption{Comparison of performances of different warm start algorithms based on the quantum approximate optimization algorithm (QAOA), as applied to the MaxCut problem, and the solution found using the Goemans-Williamson (GW) algorithm~\cite{GW} with a single random projection. The methods compared are the standard QAOA without warm start (s)~\cite{qaoa_farhi}, the WS-QAOA(WS)~\cite{wsqaoa}, the QAOA-warmest (warm)~\cite{qaoa_warmest}, the warm start adaptive-bias QAOA (WS-ab-QAOA) bias-state method (bias). (a) Difference between the calculated cut value $N_\mathrm{cut}^\mathrm{alg}$, and the exact value $N_\mathrm{cut}$, as function of the number of vertices $n$.	(b) Corresponding value of the approximation ratio $r$ as difined in Eq.~\eqref{equ:ratio}. All QAOA calculations were carried out at level-1 through the optimizer Adam~\cite{adam}, with WS-QAOA and WS-ab-QAOA initialized from the results of the GW algorithm with a single random projection and QAOA-warmest initialized from the best results of the $\mathrm{BM}$-$\mathrm{MC}_2$ algorithm running for $10$ times. Each point is the average over $40$ different unweighted-3-regular (u3r) graphs. While the QAOA, WS-QAOA and QAOA-warmest all perform worse than GW algorithm, the WS-ab-QAOA bias method is able to improve on the GW result, even at level-1.}\label{fig:dr}
\end{figure}

We first compare the calculated cut values from the warm start QAOA variants against the exact values, $N_\mathrm{cut}$, on systems with up to $180$ vertices.  The exact results come from the BiqMac method~\cite{biqmac} running on the NEOS Server~\cite{neos1,neos2,neos3}.
As shown in Fig.~\ref{fig:dr}, the various performances of the different QAOAs are compared, where the cut value differences are demonstrated in Fig.~\ref{fig:dr}(a) and approximate ratios defined by the calculated cut values divided by the exact values in Fig.~\ref{fig:dr}(b), for example, 
\begin{align}
	r_\mathrm{alg}=\frac{N_\mathrm{cut}^\mathrm{alg}}{N_\mathrm{cut}},\label{equ:ratio}
\end{align}
where $\mathrm{alg}$ represents the different algorithms. We run the GW algorithm once (do one random projection) to get a classical state and then encode it into WS-QAOA and WS-ab-QAOA as the warm start state. For the QAOA-warmest, we run the $\mathrm{BM}$-$\mathrm{MC}_2$ algorithm from $10$ different initial points and use the best result to warm start QAOA-warmest.

For the optimization of $(\gamma_1,\beta_1)$, we empirically find that $(\gamma_1=4.2315, \beta_1=1.0002)$ is a good initial point for the WS-ab-QAOA and the standard QAOA. To enhance the performance of the WS-QAOA and QAOA-warmest, the result with the maximum $\langle H_\mathrm{C} \rangle$ from $10$ different initial points is shown. It is observed that for the WS-QAOA, the initial point $(\gamma_1=5.7665,\beta_1=4.4898)$ always gives the best calculated cut values. It is clear that at level $1$, the WS-ab-QAOA gives the best approximate solution and only the WS-ab-QAOA outperforms the GW algorithm. Note that if the number of projections of the result from the first stage of the GW algorithm is increased, a better GW solution should be obtained. We will discuss this further below.

We numerically find that the calculated cut values scale linearly with $n$. For the random u3r graphs, almost all the subgraphs in the level-1 QAOA and its variants belong to the subgraph (a) in Fig.~\ref{fig:subgraph}. The optimization of the level-1 QAOA is approximately equivalent to the optimization over the subgraph (a), leading to the linearly increasing calculated cut values and parameter concentrations. For the warm start variants, a good warm start solution tends to make every edge be a cut. The optimization is roughly approximate to the optimization over the subgraph (a) with all the edges being a cut. Thus the calculated cut values are linear in $n$. It is also observed that the exact cut values are linear in $n$. As a consequence, the final approximation ratios are nearly constants in Fig.~\ref{fig:dr}(b). 

\subsection{Comparison of WS-ab-QAOA and GW algorithm}

\begin{figure}[ht]
	\centering
	\includegraphics[scale=0.55]{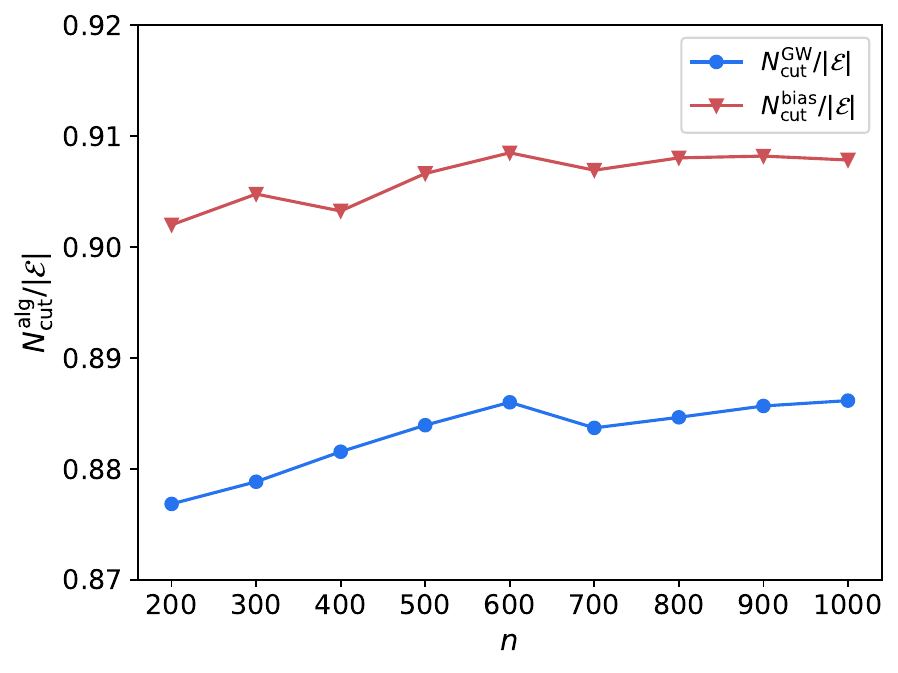} 
	\caption{Calculated cut values $N_\mathrm{cut}^\mathrm{alg}$ for the MaxCut problem of the  Goemans-Williamson (GW) algorithm $N_\mathrm{cut}^\mathrm{GW}$  (projected once)~\cite{GW} and level-1 warm start adaptive-bias quantum approximate optimization algorithm (WS-ab-QAOA) bias-state method $N_\mathrm{cut}^\mathrm{bias}$ divided by the number of edges $|\mathcal{E}|$ as a function of the number of vertices $n$. Each point is the average over $40$ different unweighted-3-regular (u3r) graphs. The classical optimizer is Adam and the starting point is $(\gamma_1=4.2315,\beta_1=1.0002)$. Again, the level-1 WS-ab-QAOA bias-state method outperforms the GW algorithm with the cut value difference about $32$ when $n=1000$. }
    \label{fig:cutn}
\end{figure}

We now turn to a direct comparison of the WS-ab-QAOA and GW algorithm. In Fig.~\ref{fig:cutn}, the system size is increased to $200\sim1000$. The exact maximal cut values are no longer available at these system sizes. Here we plot the cut values per edge for GW algorithm (projected once) and for WS-ab-QAOA. As in Fig.~\ref{fig:cutn}, the level-1 WS-ab-QAOA outperforms the GW algorithm. System size does not seem to matter much at all.

\begin{figure}[ht]
	\centering
	\includegraphics[scale=0.55]{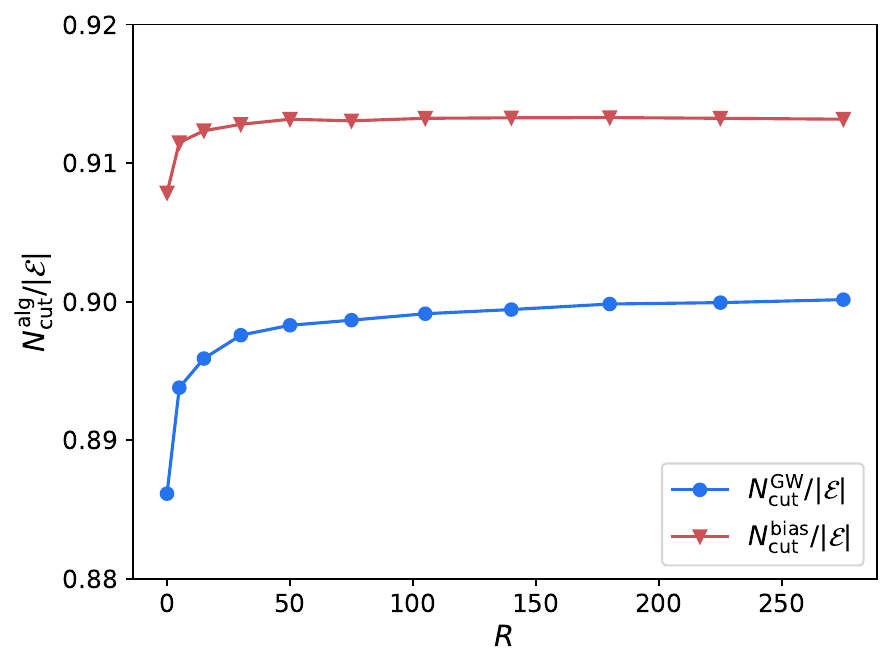} 
	\caption{Calculated cut values $N_\mathrm{cut}^\mathrm{alg}$ for the MaxCut problem of the  Goemans-Williamson (GW) algorithm $N_\mathrm{cut}^\mathrm{GW}$~\cite{GW} and level-1 warm start adaptive-bias quantum approximate optimization algorithm (WS-ab-QAOA) bias-state method $N_\mathrm{cut}^\mathrm{bias}$ divided by the number of edges $|\mathcal{E}|$ as a function of $R$, which represents the extra runs (extra projections) of the GW algorithm besides that in Fig.~\ref{fig:cutn} ($R=0$) with $n=1000$. Each point is the average over the same $40$ unweighted-3-regular (u3r) graphs as those in Fig.~\ref{fig:cutn} with $n=1000$. $R$ is set to be $5(m+1)m/2$ for m in $\{0,1,2,3,...,10\}$. The performances of the GW algorithm are slowly growing as $R$ increases, but the WS-ab-QAOA bias-state method can still improve them further.} \label{fig:gwab}
\end{figure}

We now investigate the improvement in the results on running the GW algorithm with $R+1$ different projections in Fig.~\ref{fig:gwab}, where $1$ represents the results in Fig.~\ref{fig:cutn}. As expected, the performance of GW algorithm improves as $R$ increases. However, we can use the improved solution as a starting point for the WS-ab-QAOA. We then compare performances of the improved GW solutions and the corresponding level-1 WS-ab-QAOA. This is shown in Fig.~\ref{fig:gwab} with $n=1000$. We see that the level-1 WS-ab-QAOA can still improve the GW solution further. Comparison of Figs.~\ref{fig:cutn} and~\ref{fig:gwab} makes it clear that even $275$ extra projections of the GW algorithm can not outperform the level-1 ab-QAOA warm started from the GW algorithm with only one random projection. 

\begin{figure}[ht]
	\centering
	\includegraphics[scale=0.55]{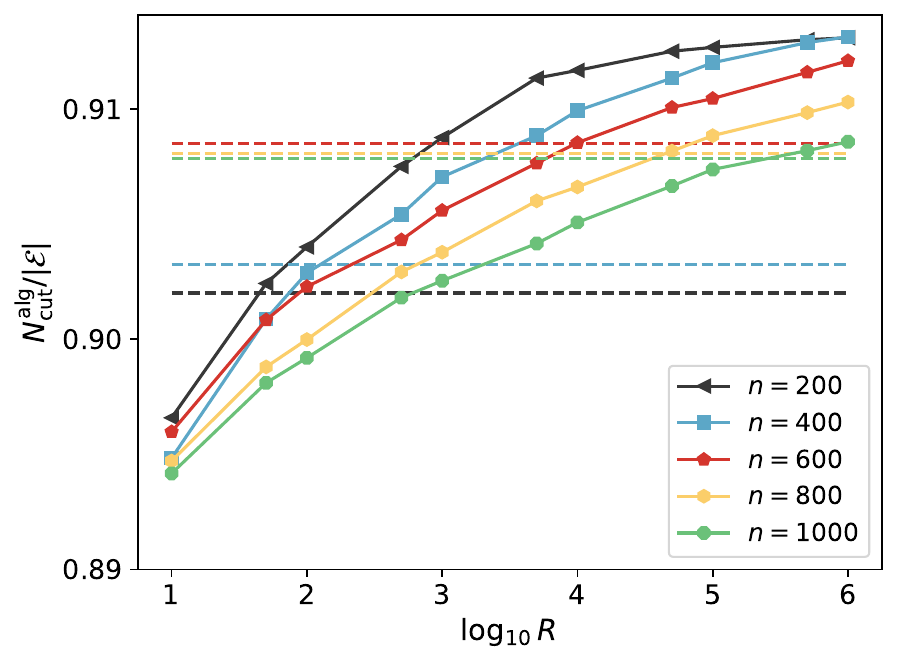} 
	\caption{Calculated cut values $N_\mathrm{cut}^\mathrm{alg}$ for the MaxCut problem of the Goemans-Williamson (GW) algorithm (solid lines)~\cite{GW} divided by the number of edges $|\mathcal{E}|$ as a function of $R$ and $n$. $R$ represents the extra runs (extra projections) of the GW algorithm besides those in Fig.~\ref{fig:cutn} ($R=0$) and $n$ means the system size of the calculated graphs. Dashed lines represent the level-1 warm start adaptive-bias quantum approximate optimization algorithm (WS-ab-QAOA) bias-state results. $R$ is set to be $\{10,50,100,500,...,10^6\}$. The same colors represent the same $n$. Each point is the average over $40$ different unweighted-3-regular (u3r) graphs. For $n=1000$, one needs about $10^{5.5}$ extra projections of the GW algorithm to outperform a single run of the ab-QAOA. The $x$ axis is on a log scale. }\label{fig:nR1}
\end{figure}

Fig.~\ref{fig:gwab} shows that a single run of the level-1 WS-ab-QAOA has the potential to outperform the GW algorithm with multiple projections. To clarify this issue we ask how many extra projections of the GW algorithm can outperform the level-1 WS-ab-QAOA warm started from a single run of the GW algorithm. We run the GW algorithm once for the warm started GW solution and then
we run the GW algorithm $R$ times for from $10 \leq R \leq 10^6$. The best cut values among these $R+1$ results are selected and shown in Fig.~\ref{fig:nR1}. Remarkably, as the system size increases it is harder and harder for the GW algorithm to obtain the same improvement as the level-1 WS-ab-QAOA. For some instances with $n=1000$, even $R=10^6$ is not enough.

\begin{figure}[ht]
	\centering
	\includegraphics[scale=0.55]{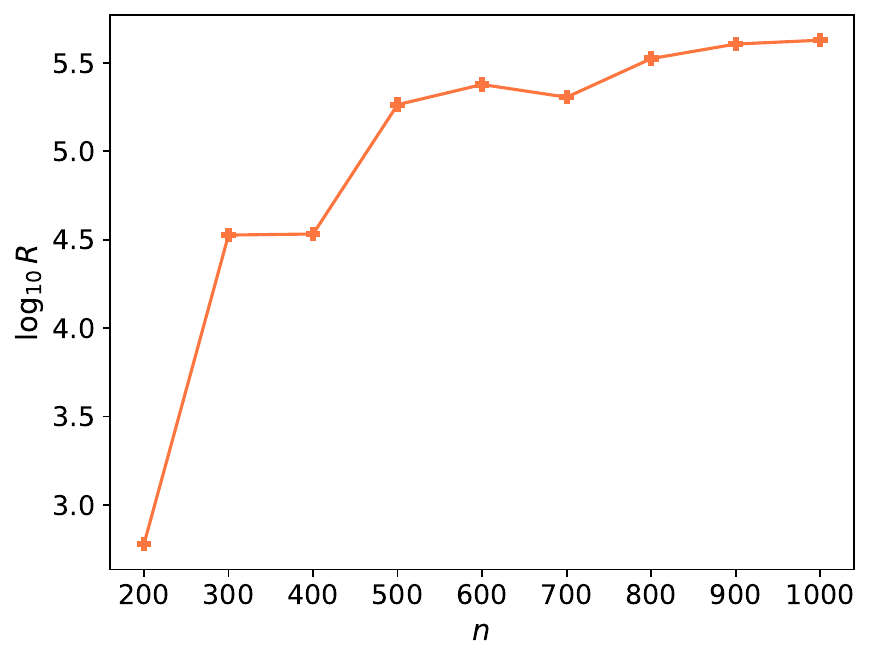} 
	\caption{Number of extra runs (extra projections) $R$ for the MaxCut problem of the Goemans-Williamson (GW) algorithm~\cite{GW} for getting a competitive result to the single run of the level-1 warm start adaptive-bias quantum approximate optimization algorithm (WS-ab-QAOA) bias-state method as a function of the system size $n$. Each point is the average over $40$ different unweighted-3-regular (u3r) graphs. The $y$ axis is in the log-scale. The improvement of the WS-ab-QAOA is more clear with $n$ increasing. $\log_{10}R$ approaches $6$ when $n>500$ means $R=10^6$ is not enough since $10^6$ is the threshold for stopping.}\label{fig:nR2}
\end{figure}

In Fig.~\ref{fig:nR2}, we compare the extra runs of the GW algorithm needed to obtain a result no worse than the level-1 WS-ab-QAOA as the speedup. The procedure is as follows. We run the GW algorithm until the GW result is no worse than the WS-ab-QAOA for the first time and record the current $R$. If the total runs exceed $10^6$, set the $R$ as $10^6$. The speedup of  level-1 WS-ab-QAOA is more clear as $n$ increases.

\section{Complexity} \label{Sec:Complexity}

For the level-1 WS-ab-QAOA on the u$\mathcal{R}$r graphs, the complexity for implementations on quantum devices is $O(n^2 \mathcal{R}^2)$, since $O(n\mathcal{R})$ quantum gates are needed to prepare $|\psi_f^\mathrm{ab} \rangle$ and $O(n\mathcal{R})$ measurements are needed to estimate $\langle H_\mathrm{C} \rangle$~\cite{abqaoa}.

For the classical simulation, with the brute-force numerical simulation method in Sec.~\ref{Sec:Simulation Method}, $\langle  H_\mathrm{C} \rangle$ can be calculated efficiently on classical computer with complexity $O(n^2\mathcal{R}^2 8^{2\mathcal{R}})$. There are $O(n\mathcal{R})$ edges and the complexity of finding the corresponding subgraphs of each edge is $O(n\mathcal{R})$, which comes from the overhead to examine all edges. Each subgraph consists of about ${2\mathcal{R}}$ qubits. The complexity of the classical representation is $2^{2\mathcal{R}}$ and the complexity of the classical matrix multiplication is $O(8^{2\mathcal{R}})$. So the total complexity for classical simulation is $O(n^2\mathcal{R}^2 8^{2\mathcal{R}})$.

With the analytical method in Sec.~\ref{Sec:Simulation Method}, the complexity of calculating one subgraph is reduced from $O(8^{2\mathcal{R}})$ to $O(2^{2\mathcal{R}-1})$, because $\tilde{R}_y(\vec{\alpha})\exp(-i\gamma_1 Z_j Z_k)\tilde{R}_y(-\vec{\alpha})$ has two different terms (see the derivation of Eq.~\eqref{equ:RyeZZ}
in Appendix~\ref{Appendix:Analytical} and this can then be applied to all $O(2\mathcal{R}-1)$ edges. The total complexity for classical simulation is $O(n^2\mathcal{R}^2 4^{\mathcal{R}})$. Although the total complexity for computing $\langle H_\mathrm{C} \rangle$ is polynomial in $n$, the level-1 WS-ab-QAOA is still a quantum algorithm, because the exact solution encoded by $|\psi_f^\mathrm{ab} \rangle$ needs to be calculated with the complexity $O(2^n)$ on a classical device. 

In the level-1 WS-ab-QAOA "bias-state method", the approximate solutions can be extracted from the bias field parameters instead of $|\psi_f^\mathrm{ab} \rangle$. The procedure is of complexity $O(n)$ on the classical device. Since the complexity for classical simulation of $\langle H_\mathrm{C} \rangle$ through the level-1 WS-ab-QAOA is $O(n^2\mathcal{R}^2 4^{\mathcal{R}})$, the total complexity including obtaining the approximate solutions in the bias-state method is $O(n^3\mathcal{R}^2 4^{\mathcal{R}})$, which is polynomial in $n$. 

The numerical results in this paper imply that a polynomial algorithm can beat the GW algorithm, which at first sight seems to be contrary to the Unique Games Conjecture that the GW algorithm yields the best possible guarantee in polynomial time~\cite{wsqaoa}. 
The point, however, is that the GW algorithm works for all graphs, while the level-1 WS-ab-QAOA bias-state method only works for u$\mathcal{R}$r graphs with smaller $\mathcal{R}$, because the final classical simulation complexity is polynomial in $n$ but exponential in $\mathcal{R}$. For u$3$r graphs, if the overhead is polynomial for the "bias-state method", then for large $\mathcal{R}$ graphs, the overhead will be exponential, and classical simulation will quickly become impossible.  That's where the speedup of the WS-ab-QAOA on real quantum machines over the GW algorithm may appear.

\section{Conclusions}
\label{Sec:Conclusions} 

We have investigated the conjecture that quantum advantages in optimization may be possible in the very last stage of the problem, where improvement by classical algorithms seems to run out. 
We did this by attempting to improve the Goemans-Williamson (GW) algorithm by using its solution as the basis for a warm start of the adaptive-bias quantum approximate optimization algorithm (ab-QAOA)~\cite{abqaoa,abqaoa_sat}. This warm start variant we call warm start ab-QAOA (WS-ab-QAOA).  

The problem instances are drawn from MaxCut on unweighted 3-regular graphs (u$3$r) with $40\sim1000$ vertices. Here the level-1 WS-ab-QAOA can recover the GW solution exactly. We presented a unified picture that can describe most of the warm start variants of the QAOA, such as the WS-ab-QAOA proposed in this paper, the WS-QAOA~\cite{wsqaoa}, the QAOA-warmest~\cite{qaoa_warmest} and the standard QAOA~\cite{qaoa_farhi}. We found analytical expressions in level 1 for the expectation value of the u$3$r MaxCut cost Hamiltonian, which accelerates the large-scale numerical simulation of those QAOA variants at level-1 case.

In the conventional ab-QAOA, the solutions are encoded in the output state $|\psi_f^\mathrm{ab}\rangle$~\cite{abqaoa,abqaoa_sat}, which can not be efficiently simulated on classical devices due to the exponentially large Hilbert space. In this paper, we found a "bias-state method" to extract the solution by the bias state constructed from the final bias field parameters $\vec{h}$, which can be done in $O(n)$ complexity. With the bias-state method, the WS-ab-QAOA can not only provide the calculated cut values but also the corresponding partition of the vertices which is not available for the other warm start QAOA variants in the classical simulation.
 
The performances of the WS-ab-QAOA and the corresponding bias-state method are compared with other warm start QAOA variants and the standard QAOA. The numerical results imply that the WS-ab-QAOA is the best and only the WS-ab-QAOA can beat the GW algorithm in the level-1 case. The improvement can only be matched by $10^{5.5}$ runs of the GW algorithm on the $1000$-vertex graphs. This of does not contradict the Unique Games Conjecture since only a special class of graphs is involved.  The complexity of the classical numerical simulation of the WS-ab-QAOA grows exponentially with the degree of the vertices. Furthermore, when restricted to 3-regular graphs, there exists a special version of the GW algorithm outperforming the original one used in this work~\cite{GW3r}. The combination of this specialized GW algorithm with WS-ab-QAOA is left for future work. 

The level-1 WS-ab-QAOA bias-state method demonstrated in this paper can be simulated efficiently on classical devices, in other words, the numerical results presented in Sec.~\ref{Sec:Numerical Results} do not directly imply a quantum advantage over the classical algorithm. Based on the numerical results in this paper and the fact that increasing the QAOA levels can improve existing performance~\cite{qaoa_farhi,qaoa_lukin,abqaoa,abqaoa_sat}, we conjecture that the real quantum advantage of the ab-QAOA might appear for large $n$, $\mathcal{R}$ and deep $p$, where classical simulation is impossible.

\textit{Note added.} Recently a similar analytical expression of $\langle H_\mathrm{C} \rangle$ in the level-1 case was presented~\cite{qaoa_analytical1}. The mixing Hamiltonian of PM-QAOA considered there has the form of Eq.~\eqref{equ:mixing_warm} but the starting state is $|\psi_0\rangle=|+\rangle^{\otimes n}$. The analytical results in Sec.~\ref{Sec:Simulation Method} and Ref.~\cite{qaoa_analytical1} can not be reduced to each other.




\acknowledgments
Yunlong Yu and Xiang-Bin Wang acknowledge National Natural Science Foundation of China Grants No.12174215 and No.12374473. Nic Shannon acknowledges the support of the Theory of Quantum Matter Unit, Okinawa Institute of Science and Technology Graduate University (OIST). We thank Chenfeng Cao for helpful discussions.

\appendix

\section{Goemans-Williamson algorithm}\label{Appendix:GW}

\subsection{Relaxed MaxCut Problem}
For a graph $G=(\mathcal{V}, \mathcal{E})$, the MaxCut problem is to find the solution of the following problem,
\begin{align}
    \begin{aligned}
        \max &\sum_{(j,k)\in \mathcal{E}} \frac{1}{2}(1-x_{j}x_{k}),\nonumber\\
        \mathrm{s.t.}&\quad x_j\cdot x_j=1 \quad \forall j \in \mathcal{V}.
    \end{aligned}
\end{align}
Here $x=(x_1,x_2,...,x_n)^T$ $x_j=1$ if $j$ is in set $\mathcal{V}_1$, and $x_j=-1$ if $j$ is in set $\mathcal{V}_2$. We then define the $n \times n$ matrix $X=xx^T$, then the MaxCut problem is equivalently formulated as
\begin{align}
    \begin{aligned}
        \max &\sum_{(j,k)\in \mathcal{E}} \frac{1}{2}(1-X_{jk}),\\
        \mathrm{s.t.}&\quad X_{jj}=1 \quad \forall j.
    \end{aligned}
\end{align}
We denote the maximum value by $N_\mathrm{cut}$, which is the true solution.

$X=xx^T$ is a symmetric rank-1 positive semi-definite matrix. Now we remove the rank-1 restriction, i.e., replace each number $x_j$ with a vector $\vec{y}_j$, then $x$ is replaced by $y=(\vec{y}_1,\vec{y}_2,...,\vec{y}_n)^T$ and the matrix $Y=yy^T$ is still a semi-definite matrix, $Y\succeq 0$, we obtain the following relaxation of MaxCut: 
\begin{align}
	\begin{split}
			\max &\sum_{(j,k)\in \mathcal{E}} \frac{1}{2}(1-Y_{jk}),\\
		\mathrm{s.t.}&\quad Y_{jj}=1 \quad \forall j,\\
		&\quad Y\succeq 0.
	\end{split}
\end{align}

Let the graph $G$ have the adjacency matrix $A$. We define entry-wise multiplication of matrices by
\begin{align}
	U \circ V=\sum_{uv}U_{uv}V_{uv}.\nonumber
\end{align}
If $V$ is symmetric then $	U \circ V=\mathrm{Tr}(UV)$. Define a matrix $\mathcal{I}$ with all entries to be $1$, then we will get the semi-definite program(SDP) formulation of MaxCut,  

\begin{align}
	\begin{split}
		\max &\quad  A \circ \frac{1}{4}(\mathcal{I}-Y),\\
		\mathrm{s.t.}&\quad Y_{jj}=1 \quad \forall j,\\
		&\quad Y\succeq 0.
	\end{split}\label{equ1}
\end{align}

Define a vector $\vec{e}$ with all elements to be one, then the Laplacian matrix $L$ on the graph $G$ is
\begin{align}
	L=\mathrm{Diag}(A\vec{e})-A,\nonumber
\end{align}
where $\mathrm{Diag}(\vec{v})$ means constructing a matrix using the vector $\vec{v}$ as the diagonal elements. The SDP formulation of MaxCut in Eq.~\eqref{equ1} is equivalent to the following problem: 
\begin{align}
	\begin{split}
		\max &\quad  \frac{1}{4} L\circ Y,\\
		\mathrm{s.t.} &\quad Y_{jj}= 1 \quad \forall j,\\
		&\quad Y\succeq 0.
	\end{split}
\end{align}

The Goemans-Williamson algorithm starts by solving this SDP to obtain the solution $Y$, or equivalently the relaxed vectors $\{\vec{y}_u\}$. Then pick a random hyperplane that passes through the origin and partition the vertices based on which side of the hyperplane the corresponding vectors are. This procedure is called random projection in the main text.

Choosing a random hyperplane is equivalent to choosing a random unit vector $\vec{r}$, since we can always find the normal vector to this hyperplane. So if $\vec{y}_j\cdot\vec{r}\geq 0$, put $j$ in set $\mathcal{V}_1$ and if $\vec{y}_j\cdot\vec{r}< 0$, put $j$ in set $\mathcal{V}_2$. The randomly chosen vector can be decomposed into $\vec{r}=\vec{r}_1 + \vec{r}_2$, where $\vec{r}_2$ is perpendicular to the plane expanded by $\vec{y}_j$ and $\vec{y}_k$ while $\vec{r}_1$ is in this plane. Then choosing a random unit vector $\vec{r}$ is equivalent to choosing a random vector $\vec{r}_1$ in this 2D plane.

In this plane, the probability that $\vec{r}_1$ cuts these two vectors $\vec{y}_j$ and $\vec{y}_k$ is equal to the probability that a random diameter perpendicular to $\vec{r}_1$ lies between these two vectors. Let $\theta$ be the angle between these two vectors. 
 Then the probability is 
 \begin{align}
    \frac{2\theta}{2\pi}=\frac{\theta}{\pi}.\nonumber
 \end{align}
Defining $\alpha$ by
\begin{align}
	\alpha=\min_{0\leq\theta\leq \pi} \frac{2\theta}{\pi (1-\cos\theta)}\approx0.878,
\end{align}
then the following inequality always holds:
\begin{align}
	\alpha\leq \frac{2\theta}{\pi (1-\cos\theta)}.\nonumber
\end{align}
This is equivalent to
\begin{align}
	\frac{\theta}{\pi} \geq \alpha \frac{1}{2}(1-\cos\theta)=\alpha \frac{1}{2}(1-\vec{y}_j\cdot\vec{y}_k).\nonumber
\end{align}
Define $N_\mathrm{cut}^\mathrm{GW}$ as the cut value from this random projection by  considering all the edges of the graph we have
\begin{align}
	N_\mathrm{cut}^\mathrm{GW}\geq\alpha N_\mathrm{cut}=0.878 N_\mathrm{cut}.
\end{align}

\subsection{Interior Point Method for MaxCut}
In this subsection we give a brief introduction of how to solve the relaxed MaxCut SDP problem with the interior-point method. More details can be found in~\cite{interior_point,iterior_point2}.
For the relaxed MaxCut problem defined by
\begin{align}
	\begin{split}
		\max &\quad  L\circ Y,\\
		\mathrm{s.t.} &\quad \mathrm{diag}(Y)=\frac{1}{4}\vec{e},\\
		&\quad Y\succeq 0,
	\end{split}\label{equ:sdp}
\end{align}
where the operator $\mathrm{diag}$ is the operation of taking the diagonal elements. Define $a=\vec{e}/4$, the relaxed problem has the dual SDP (DSDP),
\begin{align}
	\begin{split}
		\min &\quad a^T b, \\
		\mathrm{s.t.}&\quad \mathrm{Diag}(b)-L=Z,\\
		&\quad Z\succeq 0.
	\end{split}\label{equ:dsdp}
\end{align}
The goal of the interior point method is to optimize $Y$, $b$ and $Z$ to make the differences between the cost functions in Eqs.~\eqref{equ:sdp} and~\eqref{equ:dsdp} as small as possible.

The initial point can be set as~\cite{interior_point}
\begin{align}
	\begin{aligned}
		Y & =\mathrm{Diag}(a), \\
		b & =4.4\, \mathrm{abs}(L) a, \\
		Z & =\mathrm{Diag}(b)-L,
	\end{aligned}
\end{align} 
where $\mathrm{abs}(L)$ means taking the absolute value of all the elements in $L$.
Set the barrier parameter $\mu$ as 
\begin{align}
	\mu=\frac{\mathrm{Tr}(YZ)}{2n},
\end{align}
which can be solved from the associated barrier problem for DSDP as the Lagrange multiplier. The steps of $Y$, $b$ and $Z$ can be also obtained by 
\begin{align}
	\begin{aligned}
		\mathrm{\Delta} b & =\left(Z^{-1} \circ Y\right)^{-1}\left(\mu \mathrm{diag}\left(Z^{-1}\right)-a\right), \\
		\mathrm{\Delta} Z & =\mathrm{Diag}(\Delta b), \\
		\mathrm{\Delta} \tilde{Y} &=\mu Z^{-1}-Y-Z^{-1} \Delta Z Y, \\
		\mathrm{\Delta} Y & =\frac{1}{2}\left(\mathrm{\Delta} \tilde{Y}^T+\mathrm{\Delta} \tilde{Y}\right).
	\end{aligned}
\end{align}
At each iteration, we update $Y$, $b$ and $Z$ to $Y+\Delta Y$, $b+\Delta b$ and $Z+\Delta Z$, then calculate a new $\mu$ for the next iteration. This procedure implements the SDP calculation in the GW algorithm.

\section{$\mathrm{BM}$-$\mathrm{MC}_\kappa$ Algorithm}\label{Appendix:BM}
In $\mathrm{BM}$-$\mathrm{MC}_\kappa$ algorithm, each $x_j$ is relaxed to a $\kappa$-dimensional vector instead of an $n$-dimensional one in Eq.~\eqref{equ:maxcut_cost}~\cite{BM,qaoa_warmest}. Then the classical cost function is
\begin{align}
	\begin{aligned}
		\mathrm{maximize} &\sum_{(j,k)\in \mathcal{E}}\frac{1}{4}(x_j-x_k)^2,\\
		\mathrm{s.t.}\quad & ||x_j||=1,\\
		& x_j \in \mathbf{R}^\kappa.
	\end{aligned}\label{equ:maxcut_k}
\end{align} 
When $\kappa=3$, the optimization, which is usually non-convex, can be done by setting 
\begin{align}
	x_j=(\sin\theta_j\cos\phi_j,\sin\theta_j\sin\phi_j,\cos\theta_j).
\end{align}
When $\kappa=2$, set 
\begin{align}
	x_j=(\cos\theta_j,\sin\theta_j).
\end{align} 
With the help of classical optimization methods, we find a locally optimal $(\vec{\theta}^\mathrm{opt},\vec{\phi}^\mathrm{opt})$. Implementing a similar random projection produces the approximate solution. These optimal $(\vec{\theta}^\mathrm{opt},\vec{\phi}^\mathrm{opt})$ will be transferred to the Bloch sphere by Eq.~\eqref{equ:psi0_warmest} as the initial state of QAOA-warmest. 

\section{Special Case of level-1 WS-ab-QAOA}\label{Appendix:abQAOA}
Here we will consider two special cases of the WS-ab-QAOA without the update of the bias fields and show that the ab-QAOA can not benefit from the warm start state.

\subsection{$\gamma_{1}=\pi$}\label{Appendix:gamma1=pi}

We consider the special case where $\gamma_1=\gamma_1^0=\pi$. The output state of the WS-ab-QAOA is then,
\begin{align}
	|\psi_f^\mathrm{ab}\rangle&=\mathrm{e}^{-i\beta_{1}H_\mathrm{M}^\mathrm{ab}} \mathrm{e}^{-i\gamma_{1}^0 H_\mathrm{C}}|\psi_{0}^{\mathrm{ab}}\rangle.
\end{align} 
Note that $\alpha_j=\alpha_j^0=\pm\pi/6$ to encode the warm start solutions. If we denote the difference between $\beta_1$ and $\beta_1^0=\pi/2$ by $\delta\beta_1=\beta_1-\pi/2$  then we have,
\begin{align}
	|\psi_f^\mathrm{ab}\rangle&=\mathrm{e}^{-i\delta\beta_{1}H_\mathrm{M}^\mathrm{ab}}\mathrm{e}^{-i\beta_{1}^0H_\mathrm{M}^\mathrm{ab}} \mathrm{e}^{-i\gamma_{1}^0 H_\mathrm{C}}|\psi_{0}^{\mathrm{ab}}\rangle.
\end{align}

The state $\mathrm{e}^{-i\beta_{1}^0H_\mathrm{M}^\mathrm{ab}} \mathrm{e}^{-i\gamma_{1}^0 H_\mathrm{C}}|\psi_{0}^{\mathrm{ab}}\rangle$ is exactly the input classical state $|\psi_\mathrm{c}\rangle$. The expectation value of $H_\mathrm{C}$ in the state $|\psi_\mathrm{c}\rangle$, the minus calculated cut values, is given by 
\begin{align}
	\langle H_\mathrm{C} \rangle =\langle \psi_\mathrm{c}|\mathrm{e}^{i\delta\beta_{1}H_\mathrm{M}^\mathrm{ab}} H_\mathrm{C} \mathrm{e}^{-i\delta\beta_{1}H_\mathrm{M}^\mathrm{ab}} |\psi_\mathrm{c}\rangle.
\end{align}
Denote the cut values encoded by the state $|\psi_\mathrm{c}\rangle$ by ${N}_\mathrm{cut}^\mathrm{c}$   
\begin{align}
	{N}_\mathrm{cut}^\mathrm{c}=\frac{N_\mathrm{edge}}{2}-\sum_{(j,k)\in\mathcal{E}}\frac{1}{2}\langle \psi_\mathrm{c}|  Z_j Z_k|\psi_\mathrm{c}\rangle.
\end{align}
Then we can also calculate 
\begin{align}
	\begin{aligned}
		&\mathrm{e}^{i\delta\beta_{1}H_\mathrm{M}^\mathrm{ab}} H_\mathrm{C} \mathrm{e}^{-i\delta\beta_{1}H_\mathrm{M}^\mathrm{ab}}\\
        =&-\frac{N_\mathrm{edge}}{2}+\sum_{(j,k)\in\mathcal{E}}\frac{1}{2}\biggl[(1-2\cos^2\alpha_{j}^0\sin^2\delta\beta_{1})Z_j\\
        -&\cos\alpha_{j}^0\sin2\delta\beta_1 Y_j-\sin2\alpha_{j}^0\sin^2\delta\beta_1 X_j\biggr]\\
        &\biggl[(1-2\cos^2\alpha_{k}^0\sin^2\delta\beta_{1})Z_k-\cos\alpha_{k}^0\sin2\delta\beta_1 Y_k\\
        -&\sin2\alpha_{k}^0\sin^2\delta\beta_1 X_k\biggr].
	\end{aligned}
\end{align}
Only the $ZZ$ term contributes to the final expectation value since $|\psi_\mathrm{c}\rangle$ is a computational basis state, we have
\begin{align}
    \begin{split}
    	\langle H_\mathrm{C} \rangle =& -\frac{N_\mathrm{edge}}{2}+\frac{1}{2}\sum_{(j,k)\in \mathcal{E}} (1-2\cos^2\alpha_{j}^0\sin^2\delta\beta_{1})\\
        &(1-2\cos^2\alpha_{k}^0\sin^2\delta\beta_{1}) \langle \psi_\mathrm{c}|  Z_j Z_k|\psi_\mathrm{c}\rangle.
    \end{split}
\end{align}

Recall that $\alpha_j^0=\pm \pi/6$, then we have 
\begin{align}
	\begin{aligned}
		\langle H_\mathrm{C} \rangle =&\left(\frac{1+3\cos2\delta\beta_1}{4}\right)^2\sum_{jk}\frac{1}{2}  \langle \psi_\mathrm{c}|  Z_j Z_k|\psi_\mathrm{c}\rangle-\frac{N_\mathrm{edge}}{2},\\
		=&\left(\frac{1+3\cos2\delta\beta_1}{4}\right)^2(\frac{N_\mathrm{edge}}{2}-{N}_\mathrm{cut}^\mathrm{c})-\frac{N_\mathrm{edge}}{2}.
	\end{aligned}
\end{align}
What we expect is to obtain a better cut value from the WS-ab-QAOA output, which means $-\langle H_\mathrm{C} \rangle>{N}_\mathrm{cut}^\mathrm{c}$. This inequality gives
\begin{align}
	\frac{N_\mathrm{edge}}{2}>{N}_\mathrm{cut}^\mathrm{c}.
\end{align}

If we randomly put a vertex $j$ in set $\mathcal{V}_1$ or $\mathcal{V}_2$ with the probability $1/2$, then the probability of the edge $(j,k)$ being a cut is $1/2$. On average, the total number of cuts in this random algorithm is $\frac{N_\mathrm{edge}}{2}$. If we want to improve the warm start state, the warm start state should be worse than this random guess algorithm. The cut values produced by the GW-algorithm must be better than $\frac{N_\mathrm{edge}}{2}$. Thus it can not be improved. 

\subsection{Initial State is the Classical Solution}

In this appendix, we show the case where the initial state of level-1 WS-ab-QAOA is exactly the classical solution state $|\psi_\mathrm{c}\rangle$. To implement this, we set $\vec{\delta}=0$. The output state is, 
\begin{align}
    \begin{aligned}
        |\psi_f^\mathrm{ab}\rangle=\mathrm{e}^{-i\beta_{1}H_\mathrm{M}^\mathrm{ab}}\mathrm{e}^{-i\gamma_{1} H_\mathrm{C}}|\psi_\mathrm{c}\rangle.
    \end{aligned}
\end{align}
Following the analysis in Sec.~\ref{Appendix:gamma1=pi} yields
\begin{align}
	\begin{aligned}
		\langle H_\mathrm{C} \rangle &=\left(\frac{1+3\cos2\beta_1}{4}\right)^2(\frac{N_\mathrm{edge}}{2}-{N}_\mathrm{cut}^\mathrm{c})-\frac{N_\mathrm{edge}}{2},
	\end{aligned}
\end{align}
The maximum value is
\begin{align}
\max (-\langle H_\mathrm{C} \rangle)	=\left\{\begin{array}{l} 
			{N}_\mathrm{cut}^\mathrm{c},\quad\mathrm{if}\; 	{N}_\mathrm{cut}^\mathrm{c}\geq\frac{N_\mathrm{edge}}{2},\\
			\frac{15 N_\mathrm{edge}}{32}+\frac{{N}_\mathrm{cut}^\mathrm{c}}{16}<\frac{N_\mathrm{edge}}{2},\quad\mathrm{if}\; {N}_\mathrm{cut}^\mathrm{c}<\frac{N_\mathrm{edge}}{2}.
		\end{array} \right.		
\end{align}
It is obvious that there is no improvement on the GW-solution.

\section{WS-QAOA}\label{Appendix:WSQAOA}

For the level-1 WS-QAOA, if we set $\gamma_{1}=0$ and $\beta_1=\pi/2$, for the state of qubit $j$,
\begin{align}
	\begin{aligned}
			|\psi_{f}^{\mathrm{ws}} \rangle_j&=R_y^j(\pi-\theta_j)\mathrm{e}^{-i\beta_{1} X_j} R_y^j(\theta_j-\pi) R_y^j(\theta_j) |-\rangle_j\\
			&=\frac{i}{\sqrt{2}}\left(\begin{array}{c}
			-\cos\frac{3\theta_j}{2}+\sin\frac{3\theta_j}{2} \\
			\cos\frac{3\theta_j}{2}+\sin\frac{3\theta_j}{2}  
		\end{array}\right)_j\\
            &=i\left(\begin{array}{c}
			\cos\frac{3}{2}(\theta_j-\frac{\pi}{2}) \\
			-\sin\frac{3}{2}(\theta_j-\frac{\pi}{2})   
		\end{array}\right)_j,
	\end{aligned}
\end{align} 
 which reproduces the input warm start state with $\theta_j-\pi/2=\pi/3\rightarrow |1\rangle(c_j=1)$ and 
$\theta_j-\pi/2=2\pi/3\rightarrow |0\rangle(c_j=0)$. The regularization parameter $\epsilon$ can be set to be $0.25$ to reproduce the input warm start state, which means
\begin{align}
	N_\mathrm{cut}^\mathrm{WS}\geq N_\mathrm{cut}^\mathrm{GW}.
\end{align}

\section{QAOA-warmest}\label{Appendix:QAOAwarmest}

After the optimization of $\mathrm{BM}$-$\mathrm{MC}_\kappa$ algorithm, the exact relative position of different $j$ from the locally optimal $(\vec{\theta}^\mathrm{opt},\vec{\phi}^\mathrm{opt})$ is clear. However, the place of the first component $(\theta_0^\mathrm{opt},\phi_0^\mathrm{opt})$ is unclear and it plays an important role for $\langle H_\mathrm{C} \rangle$~\cite{qaoa_warmest_rotation}. There are two different strategies in Ref.~\cite{qaoa_warmest,qaoa_warmest_rotation} to decide the place,
\begin{enumerate}
\item "Vertex-at-top" strategy, where a vertex $x_j$ is uniformly sampled from $n$ vertices and rotated to $(0,0,1)$ for $\kappa=3$ and $(1,0)$ for $\kappa=2$. 
\item Uniform rotation strategy, apply a uniform rotation to all the vertices.
\end{enumerate}

As shown in Ref.~\cite{qaoa_warmest}, $\kappa=2$ warm-starts are no worse but easier than $\kappa=3$, so we will consider $\kappa=2$ case in this paper.  For the "vertex-at-top" strategy, if the vertex $x_i=(\cos\theta_i,\sin\theta_i)$ is selected to be rotated to $(1,0)$, the other $\theta_j$ should be set to $\theta_j=\theta_j-\theta_i$. For the uniform rotation strategy, all $\theta_j$ should be set to $\theta_j=\theta_j+\theta_j^\mathrm{rand}$ where $\theta_j^\mathrm{rand}$ is chosen uniformly in $[0,2\pi]$. It is also observed that the "vertex-at-top" strategy works a little better than the uniform rotation~\cite{qaoa_warmest}, thus it is adopted in this paper. 

For $\kappa=2$ QAOA-warmest from GW solution~\cite{qaoa_warmest,qaoa_treepara}, denote the cut values implied by GW algorithm by $N_\mathrm{cut}^\mathrm{GW}$, the level-$0$ QAOA-warmest will yield a cut value at least $\frac{3}{4} N_\mathrm{cut}^\mathrm{GW}$~\cite{qaoa_warmest,qaoa_treepara} with the uniform potation strategy applied. A brief proof is as follows.

As stated in Appendix~\ref{Appendix:GW}, the GW cut value is,
\begin{align}
	N_\mathrm{cut}^\mathrm{GW}=\sum_{(j,k)\in \mathcal{E}}\frac{\theta_{jk}}{\pi}=\sum_{(j,k)\in \mathcal{E}}\frac{1}{\pi}\arccos(x_j\cdot x_k).
\end{align} 
The cut value from level-$0$ QAOA-warmest is the minus expectation value of $H_\mathrm{C}$,
\begin{align}
    \begin{aligned}
    	N_\mathrm{cut}^\mathrm{warm}=&-\langle H_\mathrm{C} \rangle\\
        =&\sum_{(j,k)\in \mathcal{E}} \frac{1}{2}[1-\cos\theta_j\cos(\theta_j+\theta_{jk})]\\
        =&\sum_{(j,k)\in \mathcal{E}} \frac{1}{2}[1-\cos\theta_j\cos(\theta_j+\theta_{jk})].
    \end{aligned}
\end{align}
On average, 
    \begin{align}
    \begin{aligned}
    	N_\mathrm{cut}^\mathrm{warm}&=\frac{1}{2\pi}\int_{0}^{2\pi}\sum_{(j,k)\in \mathcal{E}} \frac{1}{2}[1-\cos\theta_j\cos(\theta_j+\theta_{jk})]\mathrm{d}\theta_j\\
        &=\sum_{(j,k)\in \mathcal{E}} \frac{1}{2}[1-\frac{1}{2}\cos(\theta_{jk})].
    \end{aligned}
\end{align}
Note that,
\begin{align}
    \begin{aligned}
        \frac{1}{2}[1-\frac{1}{2}\cos(\theta_{jk})]\geq \min_t \frac{\frac{1}{2}[1-\frac{1}{2}\cos(t)]}{ t/\pi} \frac{1}{\pi}\theta_{jk}=\frac{3}{4}\frac{1}{\pi}\theta_{jk}.
    \end{aligned}
\end{align} 
Thus we have 
\begin{align}
	N_\mathrm{cut}^\mathrm{warm}\geq\frac{3}{4}N_\mathrm{cut}^\mathrm{GW}.
\end{align}

\section{Analytical Results at low $p$.}\label{Appendix:Analytical}
In this appendix, we will present analytical results of $\langle H_\mathrm{C}\rangle$ and $\langle Z_j \rangle$. Consider two Pauli operators $P_j, P_k$. If $P_j$ commutes with $P_k$ then
\begin{align}
   \mathrm{e}^{i\phi P_j} P_k \mathrm{e}^{-i\phi P_j}	=P_k\label{equ:Pc},
\end{align}
while if $P_j$ anticommutes with $P_k$ then
\begin{align}
	\begin{split}
		&\mathrm{e}^{i\phi P_j} P_k \mathrm{e}^{-i\phi P_j}\\
		=&(\cos \phi +i\sin\phi P_j)P_k(\cos \phi -i\sin\phi P_j)\\
		=&(\cos \phi P_k +i\sin\phi P_j P_k)(\cos \phi -i\sin\phi P_j)\\
		=&\cos2\phi P_k-i\sin2\phi P_k P_j\\
		=&P_k\mathrm{e}^{-2i\phi P_j}.
	\end{split}\label{equ:Pac}
\end{align}
Recall that 
\begin{align}
	\begin{aligned}
		\langle H_\mathrm{C}\rangle
		=&\sum_{(j,k)\in \mathcal{E}}\biggl(C_j^ZC_k^Z E_{Z_jZ_k}+C_j^ZC_k^Y E_{Z_jY_k}+C_j^Y C_k^Z E_{Y_jZ_k}\\
		&+C_j^ZC_k^X E_{Z_jX_k}+C_j^X C_k^Z E_{X_jZ_k}+C_j^Y C_k^Y E_{Y_jY_k}\\
        &+C_j^X C_k^X E_{X_jX_k}+C_j^Y C_k^X E_{Y_jX_k}+C_j^X C_k^Y E_{X_jY_k}\\
        &-\frac{1}{2}\biggr),
	\end{aligned}\label{equ:EHC}
\end{align}
and 
\begin{align}
    \langle Z_j\rangle=\sum_{k\,\mathrm{s.t.}\, (j,k)\in \mathcal{E}}
		(C_j^Z E_{Z_j}+C_j^Y E_{Y_j} + C_j^X E_{X_j}).
\end{align}

We will take $E_{X_jY_k}$ as an example to show how to calculate the analytical results. The other terms can be obtained in the same way. For $E_{X_jY_k}$,
\begin{align}
	E_{X_jY_k}=\langle \psi_\mathrm{c} |\tilde{R}_y(-\vec{\delta})\mathrm{e}^{i\gamma_{1}H_\mathrm{C}}X_jY_k\mathrm{e}^{-i\gamma_{1}H_\mathrm{C}}\tilde{R}_y(\vec{\delta})|\psi_\mathrm{c}\rangle.
\end{align}
As shown in Fig.~\ref{fig:subgraph}, three different subgraphs need to be considered.

\subsection{Subgraph (a)}
At first we consider the subgraph (a) where $j_1\neq k_1, j_2\neq k_2$. In this subgraph, 
consider all the egdes, we have
\begin{align}
    \begin{aligned}
        &\mathrm{e}^{i\gamma_{1}H_\mathrm{C}}X_jY_k\mathrm{e}^{-i\gamma_{1}H_\mathrm{C}}\\
        =&\mathrm{e}^{i\gamma_{1}/2(Z_jZ_k+Z_jZ_{j_1}+Z_jZ_{j2}+Z_k Z_{k_1}+Z_k Z_{k_2})} \\
        &X_j Y_k\mathrm{e}^{-i\gamma_{1}/2(Z_jZ_k+Z_jZ_{j_1}+Z_jZ_{j2}+Z_k Z_{k_1}+Z_k Z_{k_2})}.
    \end{aligned}\label{equ:EXY}
\end{align}
Note that $Z_jZ_k$ commutes with $X_jY_k$, while the other two-body terms anticommute with $X_jY_k$. Then apply Eqs.~\eqref{equ:Pc} and~\eqref{equ:Pac} to obtain
\begin{align}
    \begin{aligned}
        &\mathrm{e}^{i\gamma_{1}H_\mathrm{C}}X_jY_k\mathrm{e}^{-i\gamma_{1}H_\mathrm{C}}\\
        =&X_j Y_k\mathrm{e}^{-i\gamma_{1}(Z_jZ_{j_1}+Z_jZ_{j2}+Z_k Z_{k_1}+Z_k Z_{k_2})}.\nonumber
    \end{aligned}
\end{align}
$E_{X_jY_k}$ can be expressed as
\begin{align}
    \begin{aligned}
    E_{X_jY_k}=&\langle \psi_\mathrm{c} |\tilde{R}_y(-\vec{\delta})X_j Y_k\\ &\mathrm{e}^{-i\gamma_{1}(Z_jZ_{j_1}+Z_jZ_{j2}+Z_k Z_{k_1}+Z_k Z_{k_2})}\tilde{R}_y(\vec{\delta})|\psi_\mathrm{c}\rangle.
    \end{aligned}
\end{align}
Inserting the identity $\tilde{R}_y(\vec{\delta})\tilde{R}_y(-\vec{\delta})$,
\begin{align}
    \begin{aligned}
    E_{X_jY_k}=&\langle \psi_\mathrm{c} |\tilde{R}_y(-\vec{\delta})X_j Y_k\tilde{R}_y(\vec{\delta})\tilde{R}_y(-\vec{\delta})\mathrm{e}^{-i\gamma_{1}Z_jZ_{j_1}}\\
    &\tilde{R}_y(\vec{\delta})\tilde{R}_y(-\vec{\delta})\mathrm{e}^{-i\gamma_1Z_jZ_{j2}}\tilde{R}_y(\vec{\delta})\tilde{R}_y(-\vec{\delta})\\
    &\mathrm{e}^{-i\gamma_1Z_k Z_{k_1}}\tilde{R}_y(\vec{\delta})\tilde{R}_y(-\vec{\delta})\mathrm{e}^{-i\gamma_1Z_k Z_{k_2}}\tilde{R}_y(\vec{\delta})|\psi_\mathrm{c}\rangle
    \end{aligned}.\nonumber
\end{align}

Note that for $\exp(i\gamma_1 Z_j Z_{j_1})$, we have,
\begin{align}
    \exp(-i\gamma_1 Z_j Z_{j_1})=\cos\gamma_1-i\sin\gamma_1 Z_j Z_{j_1},\nonumber
\end{align}
we can easily calculate
\begin{align}
    \begin{aligned}
        &\tilde{R}_y(-\vec{\delta})\mathrm{e}^{-i\gamma_{1}Z_jZ_{j_1}}\tilde{R}_y(\vec{\delta})\\
        =&\cos\gamma_{1}-i\sin\gamma_{1}(\cos\delta_{j}Z_j-\sin\delta_{j}X_j)\\&(\cos\delta_{j_1}Z_{j_1}-\sin\delta_{j_1}X_{j_1}),
    \end{aligned}\label{equ:RyeZZ}
\end{align}
and 
\begin{align}
    \tilde{R}_y(-\vec{\delta})X_jY_k\tilde{R}_y(\vec{\delta})=Y_k(\cos\delta_{j}X_j+\sin\delta_{j}Z_j).\nonumber
\end{align}
With the analytical treatment given above, we have 
\begin{align}
	\begin{split}
		E_{X_jY_k}
		=&\langle \psi_\mathrm{c} |Y_k(\cos\delta_{j}X_j+\sin\delta_{j}Z_j)\\
		&[\cos\gamma_{1}-i\sin\gamma_{1}(\cos\delta_{j}Z_j-\sin\delta_{j}X_j)\\
        &(\cos\delta_{j_1}Z_{j_1}-\sin\delta_{j_1}X_{j_1})]\\
		&[\cos\gamma_{1}-i\sin\gamma_{1}(\cos\delta_{j}Z_j-\sin\delta_{j}X_j)\\
        &(\cos\delta_{j_2}Z_{j_2}-\sin\delta_{j_2}X_{j_2})]\\
		&[\cos\gamma_{1}-i\sin\gamma_{1}(\cos\delta_{k}Z_k-\sin\delta_{k}X_k)\\
        &(\cos\delta_{k_1}Z_{k_1}-\sin\delta_{k_1}X_{k_1})]\\
		&[\cos\gamma_{1}-i\sin\gamma_{1}(\cos\delta_{k}Z_k-\sin\delta_{k}X_k)\\
        &(\cos\delta_{k_2}Z_{k_2}-\sin\delta_{k_2}X_{k_2})]|\psi_\mathrm{c}\rangle.
	\end{split}\nonumber
\end{align}
$|\psi_\mathrm{c}\rangle$ is a computational basis state so only the Pauli $Z$ term can give the non-zero result. Then the terms including $X_{j_1}$, $X_{j_2}$, $X_{k_1}$ and $X_{k_2}$ are excluded. Note that 
\begin{align}
    \begin{aligned}
        &(\cos\delta_{j}Z_j-\sin\delta_{j}X_j)^2=1,\\
        &(\cos\delta_{j}X_j+\sin\delta_{j}Z_j)(\cos\delta_{j}Z_j-\sin\delta_{j}X_j)=-i Y_j.
    \end{aligned}\nonumber
\end{align}
Thus an odd number of factors $(\cos\delta_{j}Z_j-\sin\delta_{j}X_j)$ and an even number of factors $(\cos\delta_{k}Z_k-\sin\delta_{k}X_k)$ will give zero for $E_{X_jY_k}$. As a result, we only need to consider the even numbers of $(\cos\delta_{j}Z_j-\sin\delta_{j}X_j)$ and odd numbers of $(\cos\delta_{k}Z_k-\sin\delta_{k}X_k)$. The final result is
\begin{align}
    \begin{aligned}
        E_{X_jY_k}=&\cos\gamma_{1}\sin\gamma_{1}\sin\delta_{j}\sin\delta_{k}\langle Z_jZ_k\rangle_c\\
		&(\cos^2\gamma_{1}\cos\delta_{k_2}\langle Z_{k_2} \rangle_c+\cos^2\gamma_{1}\cos\delta_{k_1}\langle Z_{k_1} \rangle_c\\
		&-\sin^2\gamma_{1}\cos\delta_{k_2}\cos\delta_{j_1}
		\cos\delta_{j_2}\langle Z_{k_2} Z_{j_1}Z_{j_2}\rangle_c\\
        &-\sin^2\gamma_{1}\cos\delta_{k_1}\cos\delta_{j_1}
		\cos\delta_{j_2}\langle Z_{k_1} Z_{j_1}Z_{j_2}\rangle_c),
    \end{aligned}\label{equ:EjkXY}
\end{align}
where $\langle Z_jZ_k\rangle_c=\langle \psi_\mathrm{c}|Z_jZ_k|\psi_\mathrm{c}\rangle=\langle Z_j\rangle_c\langle Z_k\rangle_c$.

\subsection{Subgraphs (b) and (c)}

For the subgraph (b) where $j_1 = k_1$ and $j_2\neq k_2$, we can repeat the same analysis as that in subgraph (a) to obtain $E_{X_jY_k}$. Only the terms containing $\cos\delta_{j_1}\cos\delta_{k_1}\langle Z_{j_1} Z_{k_1} \rangle_c$ can be affected by the new constraint $j_1=k_1$. When $j_1 = k_1$, it is obvious that
\begin{align}
    (\cos\delta_{j_1}Z_{j_1}-\sin\delta_{j_1}Z_{j_1})(\cos\delta_{k_1}Z_{k_1}-\sin\delta_{k_1}X_{k_1})=1,
\nonumber
\end{align}
which means the terms $\cos\delta_{j_1}\cos\delta_{k_1}\langle Z_{j_1} Z_{k_1} \rangle_c$ should be replaced by $1$ in Eq.~\eqref{equ:EjkXY},
\begin{align}
    \begin{aligned}
        E_{X_jY_k}=&\cos\gamma_{1}\sin\gamma_{1}\sin\delta_{j}\sin\delta_{k}\langle Z_jZ_k\rangle_c\\
        &(\cos^2\gamma_{1}\cos\delta_{k_2}\langle Z_{k_2} \rangle_c+\cos^2\gamma_{1}\cos\delta_{k_1}\langle Z_{k_1} \rangle\\
		&-\sin^2\gamma_{1}\cos\delta_{k_2}\cos\delta_{j_1}
		\cos\delta_{j_2}\langle Z_{k_2} Z_{j_1}Z_{j_2}\rangle_c\\
        &-\sin^2\gamma_{1}	\cos\delta_{j_2}\langle Z_{j_2}\rangle_c).
    \end{aligned}\label{equ:EjkXY1}
\end{align}

For the subgraph (c) where $j_1 = k_1$ and $j_2= k_2$,  the terms $\cos\delta_{j_2}\cos\delta_{k_2}\langle Z_{j_2} Z_{k_2} \rangle_c$ should be replaced by $1$ in Eq.~\eqref{equ:EjkXY1},
\begin{align}
    \begin{aligned}
        E_{X_jY_k}=&\cos\gamma_{1}\sin\gamma_{1}\sin\delta_{j}\sin\delta_{k}\langle Z_jZ_k\rangle_c\\
        &(\cos^2\gamma_{1}\cos\delta_{k_2}\langle Z_{k_2} \rangle_c+\cos^2\gamma_{1}\cos\delta_{k_1}\langle Z_{k_1} \rangle_c\\
		&-\sin^2\gamma_{1}\cos\delta_{j_1}
		\langle  Z_{j_1}\rangle_c-\sin^2\gamma_{1}\cos\delta_{j_2}\langle Z_{j_2}\rangle_c).
    \end{aligned}
\end{align}

\subsection{Other terms}
In the same procedure, we can obtain the analytical result for the other terms in Eq.~\eqref{equ:EHC}.

1.~For $E_{Z_jZ_k}$,
\begin{align}
    E_{Z_jZ_k}=\cos\delta_{j}\cos\delta_{k} \langle Z_jZ_k\rangle_c.\nonumber
\end{align}

2.~For $E_{Z_jY_k}$,
\begin{align}
	\begin{split}      
    E_{Z_jY_k}=&\sin\gamma_1\sin\delta_k\langle Z_k\rangle_c\\
    &\bigl(\cos^2\gamma_{1}\cos\delta_{j}\cos\delta_{k_2}\langle Z_jZ_{k_2}\rangle_c\\
		&\cos^2\gamma_{1}\cos\delta_{j}\cos\delta_{k_1}\langle Z_jZ_{k_1}\rangle_c+\cos^2\gamma_{1}\\
		&-\sin^2\gamma_{1}\cos\delta_{k_1}\cos\delta_{k_2}\langle Z_{k_1}Z_{k_2} \rangle_c\bigr).
	\end{split}\nonumber
\end{align}

3.~For $E_{Y_jZ_k}$, it can be obtained by exchanging $j$ and $k$ in $E_{Z_jY_k}$.

4.~For $E_{Z_jX_k}$,
\begin{align}
	\begin{split}
		E_{Z_jX_k}=&\sin\delta_k\langle Z_k \rangle_c\\
        &\bigl(\cos\delta_{j}\cos^3 \gamma_{1}\langle Z_j\rangle_c -\cos\delta_{k_2}\cos\gamma_{1}\sin^2\gamma_{1}\langle Z_{k_2}\rangle_c\\
        &-\cos\delta_{k_1}\cos\gamma_{1}\sin^2\gamma_{1}\langle Z_{k_1}\rangle_c\\
        &-\cos\delta_{j}\cos\delta_{k_1}\cos\delta_{k_2}\cos\gamma_{1}\sin^2\gamma_{1}\langle Z_jZ_{k_1}Z_{k_2}\rangle_c\bigr).
	\end{split}\nonumber
\end{align}

5.~For $E_{X_jZ_k}$, it can be obtained by exchanging $j$ and $k$ in $E_{Z_jX_k}$.

6.~For $E_{Y_jY_k}$,

\noindent(a).$j_1\neq k_1, j_2\neq k_2$,
\begin{align}
	\begin{split}
		E_{Y_jY_k}=&\cos^2\gamma_{1}\sin^2\gamma_{1}\sin\delta_{j}\sin\delta_{k}\langle Z_jZ_k\rangle_c\\
		&\bigl(\cos\delta_{j_2}\cos\delta_{k_1}\langle Z_{j_2}Z_{k_1}\rangle_c+\cos\delta_{j_2}\cos\delta_{k_2}\langle Z_{j_2}Z_{k_2}\rangle_c\\
        &+\cos\delta_{j_1}\cos\delta_{k_1}\langle Z_{j_1}Z_{k_1}\rangle_c\\
        &+\cos\delta_{j_1}\cos\delta_{k_2}\langle Z_{j_1}Z_{k_2}\rangle_c\bigr).
	\end{split}\nonumber
\end{align}
(b). $j_1=k_1,j_2\neq k_2$
\begin{align}
	\begin{split}
		E_{Y_jY_k}=&\cos^2\gamma_{1}\sin^2\gamma_{1}\sin\delta_{j}\sin\delta_{k}\langle Z_jZ_k\rangle_c\\
		&\bigl(\cos\delta_{j_2}\cos\delta_{k_1}\langle Z_{j_2}Z_{k_1}\rangle_c+\cos\delta_{j_2}\cos\delta_{k_2}\langle Z_{j_2}Z_{k_2}\rangle_c\\
        &+1+\cos\delta_{j_1}\cos\delta_{k_2}\langle Z_{j_1}Z_{k_2}\rangle_c\bigr).
	\end{split}\nonumber
\end{align}
(c). $j_1=k_1,j_2=k_2$
\begin{align}
	\begin{split}
		E_{Y_jY_k}
		=&\cos^2\gamma_{1}\sin^2\gamma_{1}\sin\delta_{j}\sin\delta_{k}\langle Z_jZ_k\rangle_c\\
		&\bigl(\cos\delta_{j_2}\cos\delta_{k_1}\langle+Z_{j_2}Z_{k_1}\rangle_c \\
        &+\cos\delta_{j_1}\cos\delta_{k_2}\langle Z_{j_1}Z_{k_2}\rangle_c+1+1\bigr).
	\end{split}\nonumber
\end{align}

7.~For $E_{X_jX_k}$,

\noindent(a).$j_1\neq k_1, j_2\neq k_2$,
\begin{align}
	\begin{split}
	E_{X_jX_k}=&\sin\delta_{j}\sin\delta_{k}\langle Z_j Z_k\rangle_c\\
    &\bigl(-\cos^2\gamma_{1}\sin^2\gamma_{1}\cos\delta_{k_1}\cos\delta_{k_2}\langle Z_{k_1}Z_{k_2} \rangle_c\\
		&-\cos^2\gamma_{1}\sin^2\gamma_{1}\cos\delta_{j_1}\cos\delta_{j_2}\langle Z_{j_1}Z_{j_2} \rangle_c\\
        &+\sin^4\gamma_{1}\cos\delta_{j_1}\cos\delta_{j_2}\cos\delta_{k_1}\cos\delta_{k_2}\\
		&\langle Z_{j_1}Z_{j_2}Z_{k_1}Z_{k_2} \rangle_c+\cos^4\gamma_{1}\bigr).
	\end{split}\nonumber
\end{align}
(b). $j_1=k_1,j_2\neq k_2$
\begin{align}
	\begin{split}
	E_{X_jX_k}=&\sin\delta_{j}\sin\delta_{k}\langle Z_j Z_k\rangle_c\\
    &\bigl(-\cos^2\gamma_{1}\sin^2\gamma_{1}\cos\delta_{k_1}\cos\delta_{k_2}\langle Z_{k_1}Z_{k_2} \rangle_c\\
		&-\cos^2\gamma_{1}\sin^2\gamma_{1}\cos\delta_{j_1}\cos\delta_{j_2}\langle Z_{j_1}Z_{j_2} \rangle_c\\
        &+\sin^4\gamma_{1}\cos\delta_{j_2}\cos\delta_{k_2}
		\langle Z_{j_2}Z_{k_2} \rangle_c+\cos^4\gamma_{1}\bigr).
	\end{split}\nonumber
\end{align}
(c). $j_1=k_1,j_2=k_2$
\begin{align}
	\begin{split}
		E_{X_jX_k}=&\sin\delta_{j}\sin\delta_{k}\langle Z_j Z_k\rangle_c\\
		&\bigl(\cos^4\gamma_{1}+\sin^4\gamma_{1}\\
        &-\cos^2\gamma_{1}\sin^2\gamma_{1}\cos\delta_{k_1}\cos\delta_{k_2}\langle Z_{k_1}Z_{k_2} \rangle_c\\
		&-\cos^2\gamma_{1}\sin^2\gamma_{1}\cos\delta_{j_1}\cos\delta_{j_2}\langle Z_{j_1}Z_{j_2} \rangle_c\bigr).
	\end{split}\nonumber
\end{align}

8.~For $E_{Y_jX_k}$, it can be obtained by exchanging $j$ and $k$ in $E_{X_jY_k}$.

9.~For $E_{Z_j}$,
\begin{align}
    E_{Z_j}=\cos\delta_{j}\langle Z_j \rangle_c.\nonumber
\end{align}

10.~For $E_{Y_j}$,
\begin{align}
	\begin{split}
		E_{Y_j}=&\sin\gamma_{1}\sin\delta_{j}\langle Z_j\rangle_c\\&\bigl(\cos^2\gamma_{1}\cos\delta_{j_2}\langle Z_{j_2}\rangle_c+\cos^2\gamma_{1}\cos\delta_{j_1}\langle Z_{j_1}\rangle_c\\
        &+\cos^2\gamma_{1}\cos\delta_{k}\langle Z_{k}\rangle_c\\
        &-\sin^2\gamma_{1}\cos\delta_{k}\cos\delta_{j_1}\cos\delta_{j_2}\langle Z_{k}Z_{j_1}Z_{j_2}\rangle_c\bigr).
	\end{split}\nonumber
\end{align}

11.~For $E_{X_j}$,
\begin{align}
	\begin{split}
		E_{X_j}=&\cos\gamma_{1}\sin\delta_{j}\langle Z_j \rangle_c\\
        &\bigl(\cos^2\gamma_{1}-\sin^2\gamma_{1}\cos\delta_{j_1}\cos\delta_{j_2}\langle Z_{j_1}Z_{j_2} \rangle_c\\
		&-\sin^2\gamma_{1}\cos\delta_{j_1}\cos\delta_{k}\langle Z_{j_1}Z_{k}\rangle_c\\
         &-\sin^2\gamma_{1}\cos\delta_{k}\cos\delta_{j_2}\langle Z_{k}Z_{j_2}\rangle_c\bigr). 
	\end{split}\nonumber
\end{align}

\section{Additional Results}\label{Sec:diff}

\begin{figure}[ht]
	\centering
	\includegraphics[scale=0.55]{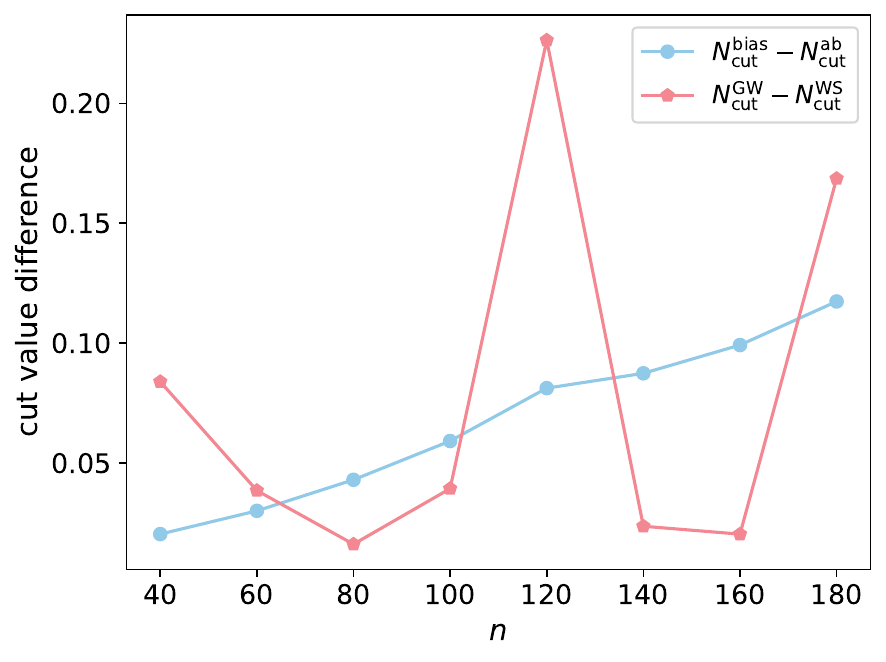} 
	\caption{Cut value differences for the MaxCut problem between the level-1 warm start quantum approximate optimization algorithm (WS-ab-QAOA) $N_\mathrm{cut}^\mathrm{ab}$ and the corresponding bias-state method $N_\mathrm{cut}^\mathrm{bias}$ and between the Goemans-Williamson (GW) algorithm~\cite{GW} $N_\mathrm{cut}^\mathrm{GW}$ and the corresponding WS-QAOA~\cite{wsqaoa} $N_\mathrm{cut}^\mathrm{WS}$. There is little difference between the $N_\mathrm{cut}^\mathrm{ab}$ and $N_\mathrm{cut}^\mathrm{bias}$, implying the convergence of WS-ab-QAOA to the computational basis state. The difference between $N_\mathrm{cut}^\mathrm{GW}$ and $N_\mathrm{cut}^\mathrm{WS}$ is also negligible.}\label{fig:d}
\end{figure}

Fig.~\ref{fig:d} shows that there is very little difference ($ < 0.1$) between $N_\mathrm{cut}^\mathrm{bias}$ and $N_\mathrm{cut}^\mathrm{ab}$. This means that $|\psi_{f}^{\mathrm{ab}}\rangle$ converges to the bias state $|\psi_{\mathrm{bias}}\rangle$. This justifies us in only considering $N_\mathrm{cut}^\mathrm{bias}$ in the main text. There is also very small difference between $N_\mathrm{cut}^\mathrm{GW}$ and $N_\mathrm{cut}^\mathrm{WS}$. This implies that the input GW solution seems to be the global optimum of level-1 WS-QAOA.
 
\bibliography{ref}

\end{document}